\documentclass{jfm}
\usepackage{graphicx}
\usepackage{color}
\usepackage{epstopdf, epsfig}
\usepackage{latexsym,bm}
\usepackage{longtable}

\usepackage{hyperref}
\hypersetup{
    colorlinks=true,
    linkcolor=blue,
    filecolor=blue,      
    urlcolor=blue,
    citecolor=blue,
}

\shorttitle{Spherical Rotating Convection}
\shortauthor{Y. Lin and A. Jackson}

\title{Large-scale vortices and zonal flows in spherical rotating convection}

\author{Yufeng Lin\aff{1}
  \corresp{\email{linyf@sustech.edu.cn}},
\and Andrew Jackson\aff{2}}

\affiliation{\aff{1} Department of Earth and Space Sciences, Southern University of Science and Technology, Shenzhen 518055, China
\aff{2} Institute of Geophysics, ETH Zurich, Zurich 8092, Switzerland}

\begin{document}
\maketitle

\begin{abstract}
Motivated by understanding the dynamics of stellar and planetary interiors, we have performed a set of direct numerical simulations of Boussinesq convection in a rotating full sphere. The domain is internally heated with fixed temperature and stress-free boundary conditions, but fixed heat flux and no-slip boundary conditions are also briefly considered. We particularly focus on the large-scale coherent structures and the mean zonal flows that can develop in the system. At Prandtl number of unity, as the thermal forcing (measured by the Rayleigh number) is increased above the value for the onset of convection, we find a relaxation oscillation regime, followed by a geostrophic turbulence regime. Beyond this we see for the first time the existence of large-scale coherent vortices that form on the rotation axis. All regime boundaries are well described by critical values of the convective Rossby number $Ro_c$, with transitions from oscillatory to geostrophic turbulence, and then to the large-scale vortex regime at values $Ro_c\approx 0.2$ and $Ro_c\approx 1.5$, respectively. The zonal flow is  controlled by the convective Rossby number and changes its direction when the flow transitions from the geostrophic turbulence regime to the large-scale vortex regime. While the non-zonal flow speed and heat transfer can be described by the so-called inertial scaling in the geostrophic turbulence regime, the formation of large-scale vortices appears to reduce both the non-zonal flow speed and the efficiency of convective heat transfer. 
\end{abstract}

\section{Introduction}\label{sec:Introduction}
Convection is commonplace in the interiors of  stars and planets, in which rotation plays a dominant role \citep[e.g.][]{Zhang2017}. Despite a concerted study of this system, both numerically and theoretically, over many decades, our understanding of the process remains incomplete. It is of great interest to properly characterise both the morphology of fluid flow and the mechanism of heat transfer as a function of the dynamical regime as controlled by the relevant non-dimensional numbers. The highly supercritical regime, which is both the most challenging and the most relevant in planetary settings, 
remains poorly understood.

Three classes of problem have been tackled in order to make progress. By far the most studied is the problem of plane layer Rayleigh-B\'enard convection, in both non-rotating and rotating settings \citep{PlumleyJulien2019}. The second class is rotating spherical shell convection, naturally motivated by the geometry of planets that possess solid cores; such a geometry lends itself naturally to efficient numerical methods \citep{Glatzmaier_2014,Martigcubed2016}. This problem has been studied using laboratory experiments \citep{Aubert2001} and numerical simulations in both Boussinesq \citep{Christensen2002,Aurnou2007} and anelastic settings \citep{Gastine2013}, with 
a variety of gravity profiles \citep{Gastine2016} and boundary conditions \citep{Mound2017,Long2020}. The results of the first two problems have been recently reviewed by \cite{Jones2015,Aurnou2015} and \cite{PlumleyJulien2019}, and we refer the reader to these review articles.

The third class of study, which is the subject of the present research, is the canonical problem of rotating convection in a whole sphere in which an internal heat source is present; the strength of the heating is measured by the Rayleigh number $Ra$.
Although early theoretical work concentrated on the whole-sphere geometry \citep{Book_Chandra_61,Roberts1968,Busse1970,Kida1994,Jones2000,Zhang2004} and the determination of the critical Rayleigh number for convection ($Ra_c$), there has been much less numerical work focusing on this area. Notwithstanding the studies that reduce the size of the inner core to negligible size without actually removing it,   
notable studies using a true full-sphere geometry are those of \cite{Jones2000,Sanchez2016} and \citet{Kaplan2017}, which mainly concerned themselves with the onset of convection or subcritical convection. More recently,  \citet{Guervilly2019} examined the convective length scales in a rotating sphere using both quasi-geostrophic and fully three-dimensional methods in the regime of very low Ekman number and  small Prandtl number, albeit with a weak thermal forcing (around {the onset of convection}). Convection was studied experimentally in the full sphere 
by \citet{Chamberlain1986}  using centrifugal gravity, though practical considerations precluded the exploration of very supercritical Rayleigh numbers: in that study,  $Ra\le  2.5Ra_c$. The highly supercritical regime of convection in a whole sphere is of great relevance to the dynamics of the Earth's core prior to inner core nucleation and to some fully convective stars, yet the problem remains largely unexplored.

In the present work we concentrate on the problem in which the diffusivities of heat and momentum are equal, leading to a so-called Prandtl number ($Pr$) of unity. At fixed viscosity this leaves the {rotation period} (measured by the Ekman number $E$) and the strength of internal heating (characterised by $Ra$) as the only control parameters. We perform a set of direct numerical simulations with moderate Ekman numbers ($E\sim10^{-5}$), but over a wide range of Rayleigh numbers (from near the onset up to $10^4Ra_c$).

A primary focus of our work has been to delineate the regime boundaries that occur on increasing the Rayleigh number. We make use of the significant step forward  made by \citet{Gilman1977} who identified the relevance of the convective Rossby number $Ro_c$ in supercritical convection. In terms of the aforementioned parameters, $Ro_c=\sqrt{RaE/Pr}$, 
and we find that $Ro_c$ is indeed
the control parameter that defines the regime boundaries. { A relaxation oscillation} regime  occurs just beyond the onset of convection and the boundary with the {geostrophic turbulence} (GT) regime that follows subsequently is well described by a critical value of $Ro_c$. 
Likewise, we find a second bifurcation to a regime with a large-scale vortex
(LSV), which is perfectly described by a constant value of $Ro_c$.
Thus our results support the significance of this parameter. Conversely, when looking quantitatively within these regimes, we find that this parameter alone is insufficient to fully describe the scaling of flow speed and heat transfer. In particular, we find that the formation of LSVs appears to reduce the efficiency of the {heat} transfer and the convective flow speed.

The identification of axisymmetric LSVs in spherical rotating convection is entirely new. We argue that the omission of an inner core removes the obstacle to such geostrophic structures that otherwise would be present.
Large-scale coherent vortices have been previously observed in planar rotating convection in both the anelastic case \citep{Chan2007, Chan2013,Kapyla2011} and  Boussinesq case \citep{Julien2012GAFD,Favier2014,Guervilly2014,Stellmach2014,Rubio2014}. The formation of LSV is attributed to an inverse cascade, i.e. upscale energy transfer, in rotating turbulence. In spherical containers, large-scale coherent structures may be associated with  mean zonal flows \citep{Christensen2002}. We find a reversal
of the mean zonal flow when the LSVs are formed. Such a reversal has been observed in thin spherical shells for both  Boussinesq convection \citep{Aurnou2007} and anelastic convection \citep{Gastine2013} when the Rayleigh number is sufficiently large, but never before in a full sphere. Effects of boundary conditions on the formation of LSVs are also briefly discussed.

\section{Numerical Models}\label{sec:Numerics}
\subsection{Governing equations}
We consider Boussinesq convection in a full sphere with radius of $r_o$, which rotates at $\bm \Omega=\Omega \bm{\hat z}$. The sphere is filled with an incompressible fluid of density $\rho$, viscosity $\nu$  and thermal diffusivity $\kappa$. Convective motions are driven by a homogeneous internal heat source $S$, under a gravitational field $\bm g=-g_o\bm r/r_o$. In the absence of convection, the basic state temperature is determined by the conduction alone and is given as
\begin{equation}
{
\widetilde{T_b}}=\frac{\beta}{2}(r_o^2-r^2),
\end{equation}
where $\beta=S/(3 k)$ in which $k$ is thermal conductivity.

Using the radius $r_o$ as the length scale, the viscous diffusion time $r_o^2/\nu$ as the time scale and $\beta r_o^2$ as the temperature scale, the  system is governed by the following dimensionless equations in the rotating frame \citep{Marti2014}:
\begin{equation} \label{eq:NS}
E \left(\frac{ \partial \bm u}{\partial t}+\bm {u\cdot \nabla u}\right)+\bm{\hat z}\times \bm u=-\bm \nabla p +\frac{Ra}{Pr}T\bm r+E\nabla^2 \bm u,
\end{equation}
\begin{equation}\label{eq:Temp}
Pr \left(\frac{ \partial T}{\partial t}+\bm{u\cdot \nabla} T \right)=\nabla^2 (T-T_b),
\end{equation}
\begin{equation} \label{eq:divu}
\bm{\nabla \cdot u}=0,
\end{equation}
where $\bm u$ is the fluid velocity, $p$ is the reduced pressure, $T$ is the total temperature and $T_b$ is the dimensionless basic temperature {given} as
\begin{equation}
T_b=\frac{1}{2}(1-r^2).
\end{equation}

The dimensionless parameters are the Ekman number $E$, the rotationally modified Rayleigh number $Ra$ and the Prandtl number $Pr$:
\begin{equation}
E=\frac{\nu}{2\Omega r_o^2}, \quad Ra=\frac{\alpha \beta g_o r_o^3}{2 \Omega \kappa},\quad \Pran=\frac{\nu}{\kappa},
\end{equation}
where $\alpha$ is the thermal expansion coefficient. The Prandtl number is fixed to be unity ($\Pran=1$) in this study. Note that $Ra$ in a full sphere is naturally flux-based as $\beta$ is related to the total heat flux $\tilde{Q}$ through the boundary by 
\begin{equation}
    \tilde{Q}=\frac{4\pi r_0^3}{3}S=4 \pi r_0^3 k \beta, 
\end{equation}
regardless of  thermal boundary conditions, and the non-dimensional heat flux $q$ through the boundary is $q=-\nabla T$. 
{ Here $Ra$ is related to the convectional Rayleigh number $Ra_T=\alpha g_0\Delta Tr_o^3/\nu\kappa$ based on the temperature difference by $Ra=2NuERa_T$, where $Nu$ is the Nusselt number (see the definition given in equation (\ref{def:Nusselt}) below).}

As pointed out by \citet{Gilman1977}, the convective Rossby number, which is a combination of $E$, $Ra$ and $Pr$, is a key control parameter in supercritical rotating convection. Based on our definition, the convective Rossby number is given as
\begin{equation}
    Ro_c=\sqrt{\frac{ERa}{Pr}}=\frac{\sqrt{\alpha \beta g_0 r_o}}{2\Omega},
\end{equation}
which does not depend on the viscosity $\nu$ and the thermal diffusivity $\kappa$. { The convective Rossby number is the ratio between the rotation time scale and the free-fall time scale, and characterises the global force balance between the buoyancy force with respect to the Coriolis force \citep{Gilman1977}.}
\cite{Christensen2002}  introduced a similar diffusivity-free parameter $Ra^*=ERa/Pr$, which is the square of the convective Rossby number, i.e. $Ra^*=Ro_c^2$. 

 We adopt the stress-free boundary condition for the velocity $\bm u$ and a fixed temperature ($T=0$ for simplicity) at the boundary of the sphere in most of the numerical simulations. The no-slip boundary condition and a fixed heat flux, i.e. $\partial T/\partial r=-1.0$, boundary condition are also briefly considered. 
 
 \subsection{Numerical method}
  The governing equations (\ref{eq:NS}-\ref{eq:divu}) subject to the boundary conditions are numerically solved using a spectral method in a whole sphere \citep{Marti2016}. As we consider an incompressible fluid,  the velocity $\bm u$ is decomposed into  toroidal and poloidal components:
  \begin{equation}
      \bm u= \bm {\nabla \times}(\mathcal{T} \bm r) +\bm {\nabla \times\nabla \times}(\mathcal{P} \bm r).
  \end{equation}
  The toroidal $\mathcal{T}$ and poloidal $\mathcal{P}$ scalar fields and the temperature field $T$ are expanded in terms of spherical harmonic expansion on spherical surfaces and the so-called Jones-Worland polynomials in the radial direction. The spectral expansion is truncated up to spherical harmonics of degree $L=255$ and order $M=255$, and up to $N=127$ for the Jones-Worland polynomials.
   A critical element of the technique is its graceful handling of the origin at the centre of the sphere, such that the Courant–Friedrichs–Lewy condition is not unduly stringent in this region.
  Time-stepping has been performed using a second-order Runge-Kutta scheme, with adaptive time steps. The typical time step is about $10^{-7}$ (viscous time scale) for the most demanding calculations. We refer the reader to \cite{Marti2016} for more details regarding the numerical scheme. The numerical code has been benchmarked for several hydrodynamic and magnetohydrodynamic problems in a full sphere, including  the rotating thermal convection problem \citep{Marti2014}. 
  
\subsection{Diagnostics}
In order to qualitatively describe numerical results, the following quantities derived from the velocity $\mathbf u$ and the temperature $T$ will be used. 

The total kinetic energy is given by
\begin{equation}
\mathcal{E} =\frac{1}{2}\int_v \bm u^2 \mathrm{d}V.
\end{equation}The development of strong zonal flow is a distinct feature in the nonlinear regime, hence we can decompose the total energy into zonal and non-zonal parts:
\begin{equation}
\mathcal{E}_{\mathrm{zonal}} =\frac{1}{2}\int_v \left( u_{\phi}^0\right)^2 \mathrm{d}V, \quad \mathcal{E}_{\mathrm{non}}=\mathcal{E}-\mathcal{E}_{\mathrm{zonal}},
\end{equation}
where $ u_{\phi}^0$ is the axisymmetric ($m=0$) component of the azimuthal velocity $u_\phi$, i.e. zonal flow. 
We define the Rossby number as
\begin{equation}
Ro=\frac{U_{rms}}{2\Omega r_0},
\end{equation}
where $U_{rms}$ is the dimensional root mean square velocity. Base on the non-dimensionalisation we used, the Rossby number is related to the non-dimensional energy as
\begin{equation}
Ro=E\sqrt{\frac{2\mathcal{E}}{V}},
\end{equation}where $V=4 \pi/3$ is the non-dimensional volume of the sphere. Accordingly, we have the zonal Rossby number 
\begin{equation}
Ro_{\mathrm{zon}}=E\sqrt{\frac{2\mathcal{E_{\mathrm{zonal}}}}{V}},
\end{equation}
and the non-zonal Rossby number
\begin{equation}
Ro_{\mathrm{non}}=E\sqrt{\frac{2\mathcal{E_{\mathrm{non}}}}{V}}.
\end{equation}

Heat transport owing to convection can be measured by the Nusselt number. For the convection of internal heating in a full sphere, the Nusselt number can be estimated by the drop of the temperature with respect to the basic state at the centre \citep{Guervilly2016,Kaplan2017}:
 \begin{equation}
Nu=\frac{T_b(r=0)}{T(r=0)}.\label{def:Nusselt}
\end{equation}

Apart from these global quantities, we often use the axial vorticity and the zonal velocity to visualise flows. 
The axial vorticity is calculated by
\begin{equation}
\omega_z=2 E \mathbf{\hat{z}}\mathbf{\cdot} (\mathbf {\nabla \times u}).
\end{equation}
Note that the prefactor of $2E$ leads to the vorticity in the units of the rotation frequency $\Omega$ given the non-dimensionalisation we have used. The zonal velocity is also rescaled, i.e. $U_\phi^0=2E u_\phi^0$, such that $U_\phi^0$ is in the units of $\Omega r_0$. The mean zonal flow $\overline{U_\phi^0}$ is time-averaged $U_\phi^0$.

\section{Results}\label{sec:Results} 
\begin{figure}
\begin{center}
\includegraphics[width=0.7 \textwidth]{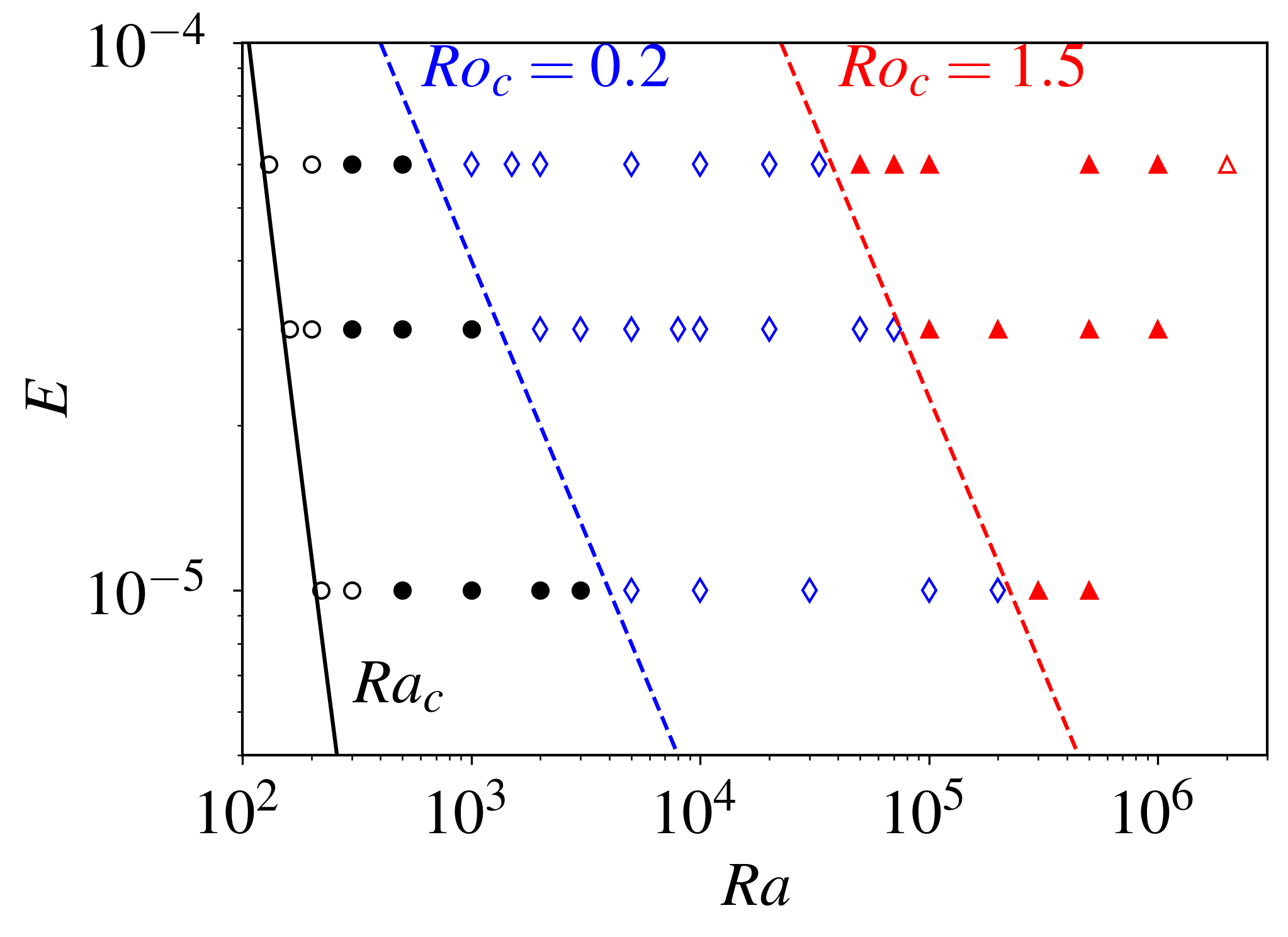}
\caption{Regime diagram in the plane of $(E, Ra)$.  Different symbols represent different flow regimes. Open circles correspond to the {steadily drifting} regime near the onset; filled circles correspond to the { relaxation  oscillation} regime; blue diamonds correspond to the GT regime; filled triangles correspond to the LSV regime; the open triangle in the top right corresponds to the non-rotating regime.}
\label{fig:RegimDiagram}
\end{center}
\end{figure}

Numerical simulations in this study are listed in detail in table \ref{tab:summary} in Appendix \ref{appA}. We fix $\Pran=1$ in this study and vary the control parameters $E$ and $Ra$, leading to a variety of dynamical regimes. {Stress-free} and fixed temperature boundary conditions are assumed unless stated otherwise. 
Figure \ref{fig:RegimDiagram} summarises simulations as a function of $(E, Ra)$ with symbols indicating different flow regimes, which are discussed in section \ref{sec:regime}.

\subsection{Flow regimes} \label{sec:regime}

\begin{figure}
\includegraphics[width=0.7 \textwidth]{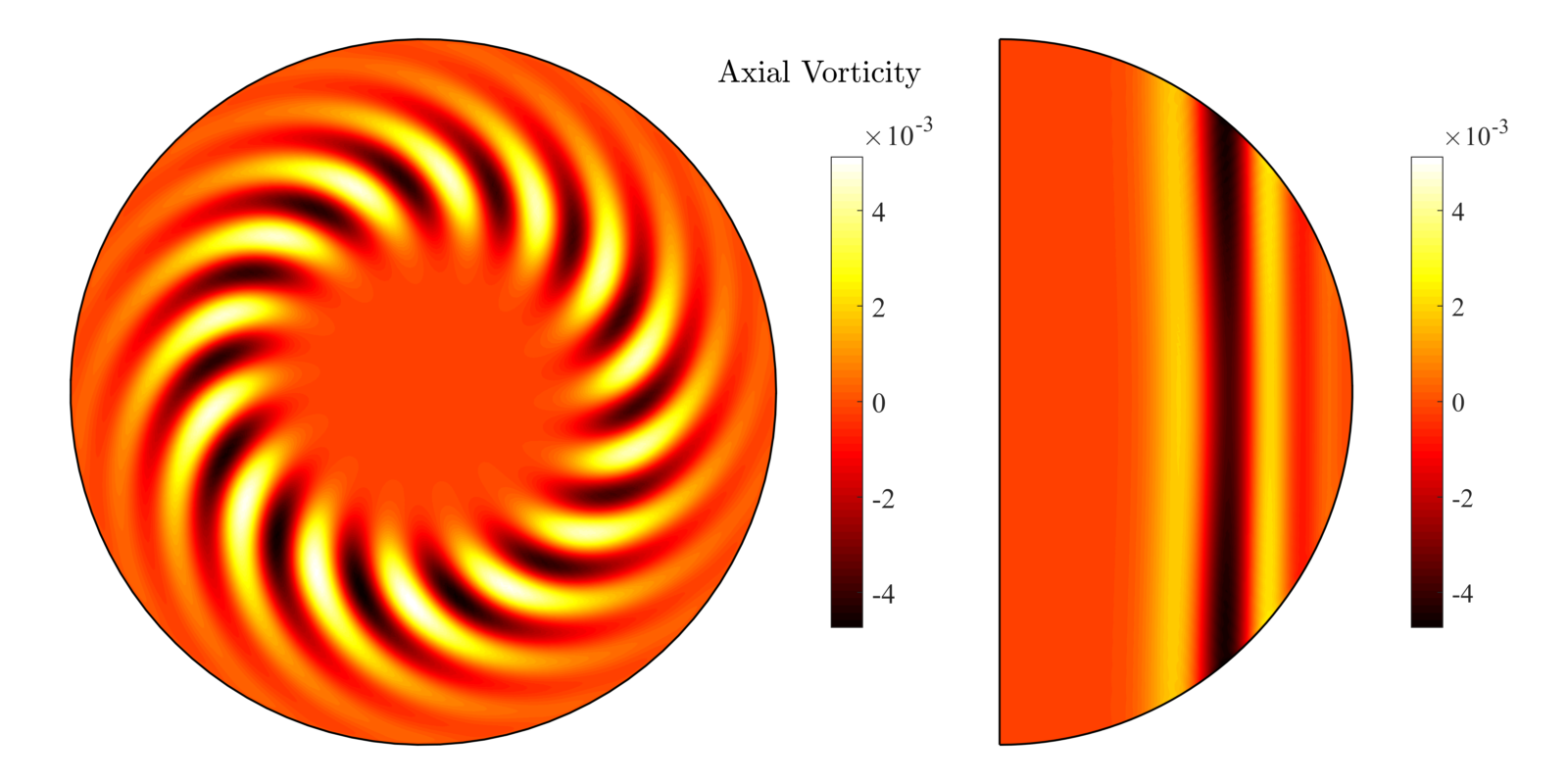}
\includegraphics[width=0.25 \textwidth]{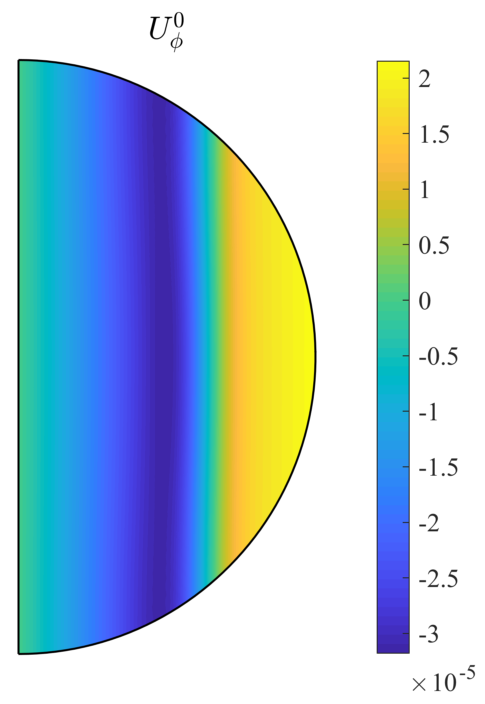} \\
\hspace*{2.5cm} (a) \hspace*{3cm} (b)  \hspace*{3 cm} (c) 
\caption{Flows near the onset at $E=10^{-5}$, $Pr=1$, $Ra=220\approx1.05Ra_c$. (a) Axial vorticity in the equatorial plane and (b) in the meridional plane; (c) zonal flow in the meridional plane. }
\label{fig:onset}
\end{figure}
Convection takes place only when $Ra$ is {larger} than a critical value $Ra_c$. At $Pr=1$ with the stress-free and the fixed temperature boundary conditions, the critical value is given as $Ra_c\approx4.1 E^{-1/3}+17.8$ (black solid line in figure \ref{fig:RegimDiagram}) according to the asymptotic theory of the linear onset \cite{Jones2000}.  Our fully nonlinear numerical simulations are consistent with the linear theory regarding the onset of convection. Figure \ref{fig:onset} shows a snapshot of the axial vorticity and the zonal flow when $Ra$ is marginally above the critical value ($Ra=220\approx 1.05Ra_c$) at $E=10^{-5}$. The flow near the onset is in the form of the so-called thermal Rossby waves, which are quasi-geostrophic, i.e. nearly invariant along the rotation axis (figure \ref{fig:onset} (b)). The kinetic energy becomes quasi-steady after the initial growing phase. The azimuthal wavenumber $m=14$ is also in agreement with the linear theory.  We also see that a weak zonal flow develops (figure \ref{fig:onset} (c)) with differential rotation (prograde near the equator and retrograde in the inner region)  due to the nonlinear interactions of the linear onset mode \citep{Zhang1992}.

\begin{figure}
\begin{center}
\includegraphics[width=0.95 \textwidth]{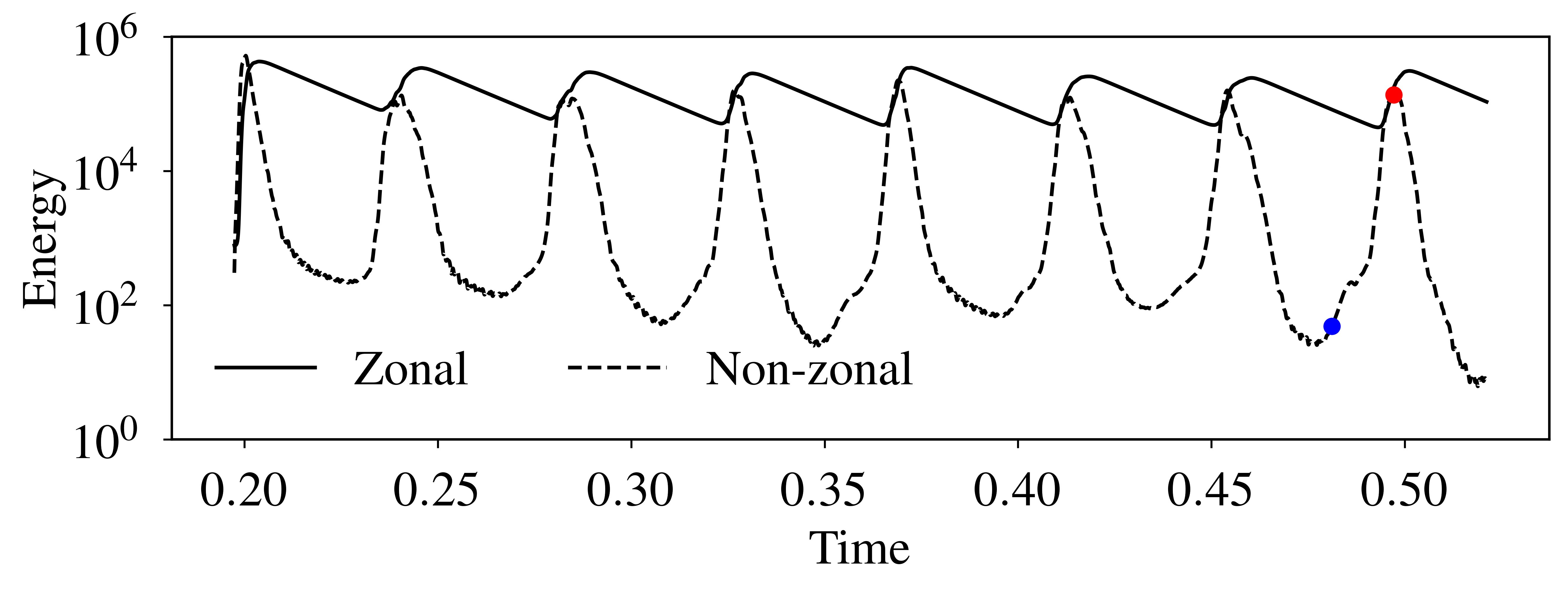} \\
(a) \\
\includegraphics[width=0.49 \textwidth]{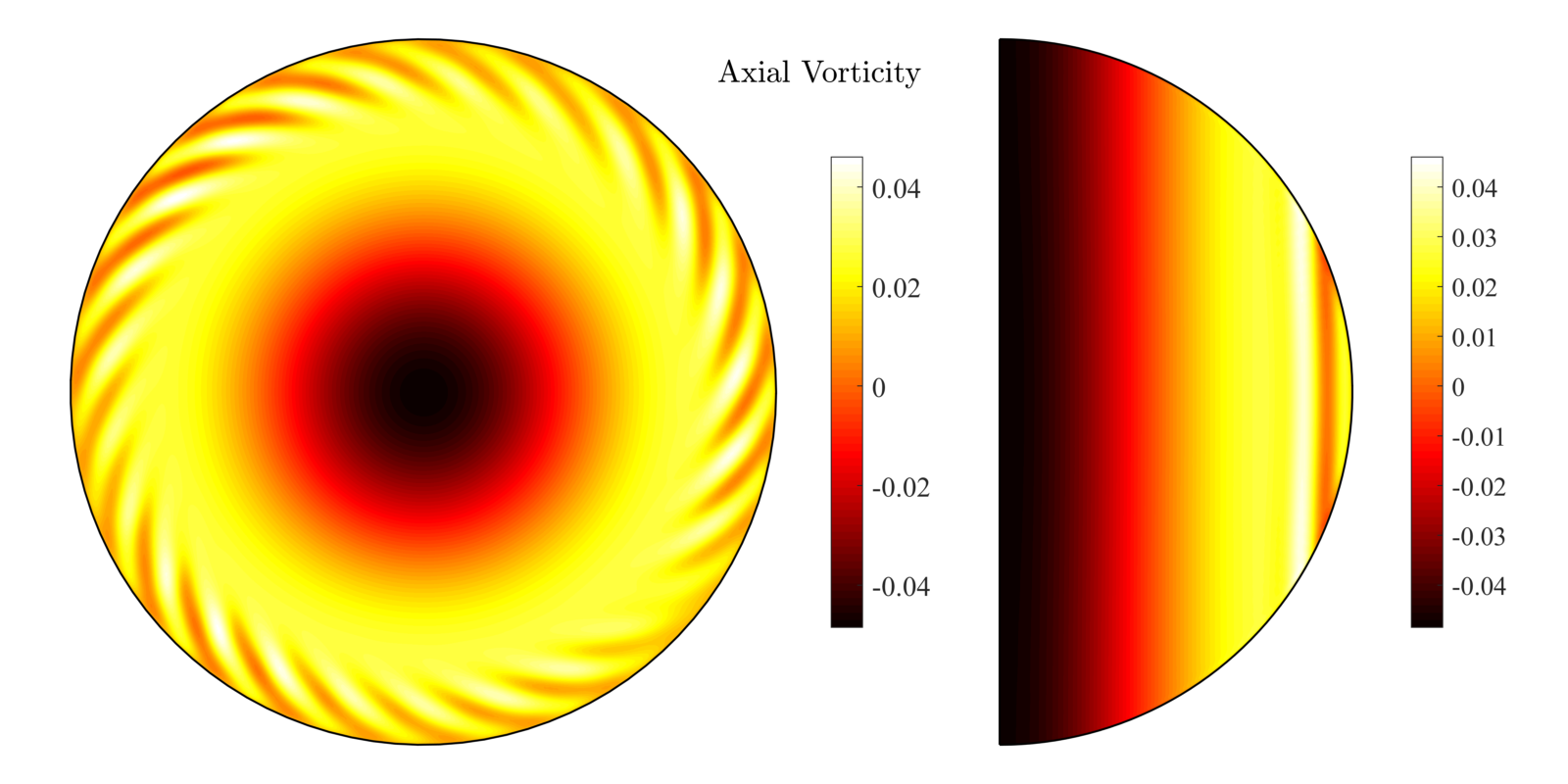}
\includegraphics[width=0.49 \textwidth]{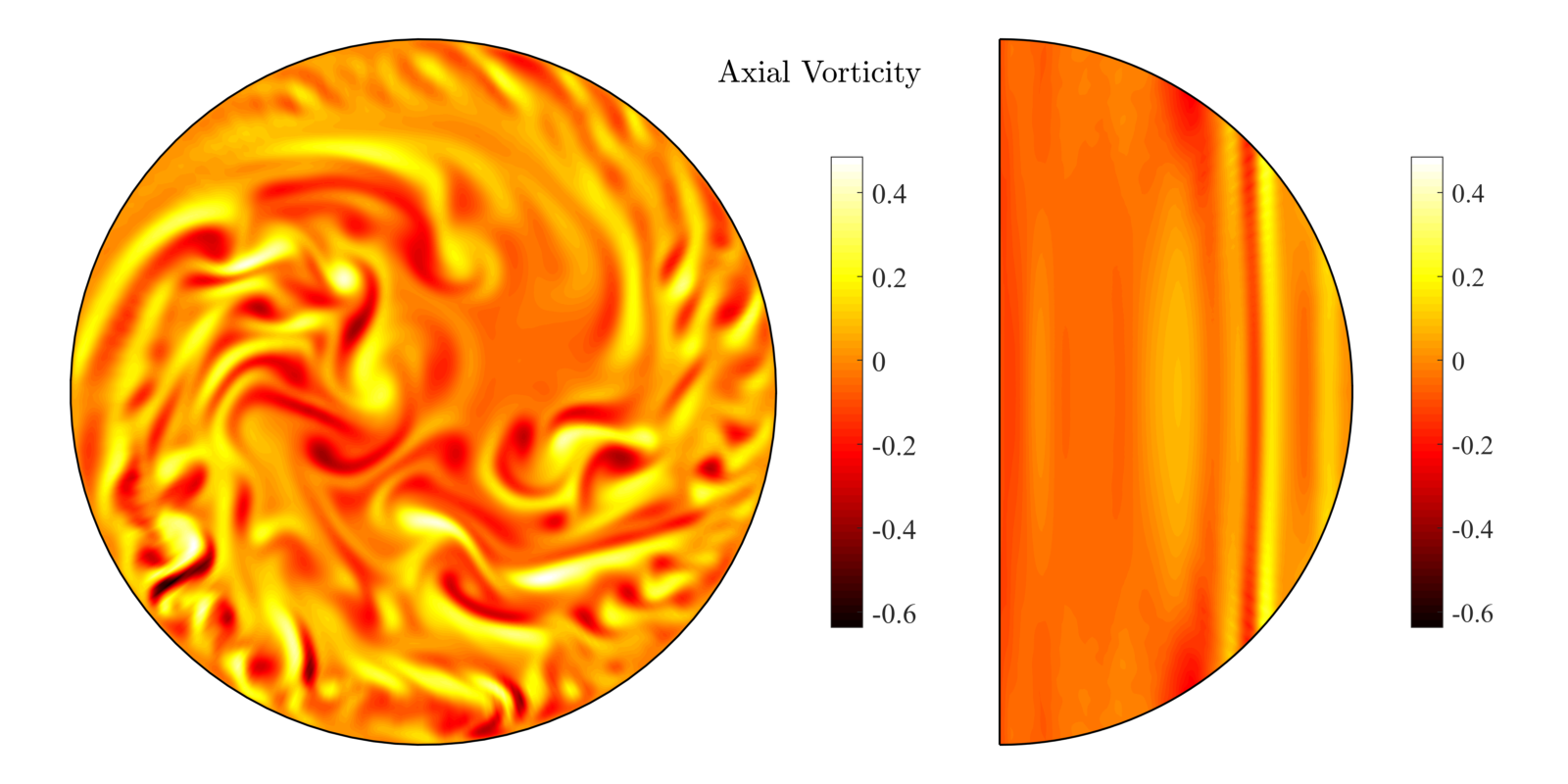} \\
(b) \hspace*{6cm} (c) \\
\caption{An example of the oscillatory flow regime at $E=10^{-5}$, $Ra=1000\approx 4.8 Ra_c$. (a) Time evolution of the zonal and non-zonal kinetic energy. Time is in units of the viscous time scale. (b)  and (c) Axial vorticity in the equatorial plane and in the meridional plane at the instants indicated by the blue dot and red dot in (a) respectively. (A movie  of this time series is provided; see \href{https://faculty.sustech.edu.cn/wp-content/uploads/2020/12/2020122222131257.mp4}{Movie 1}).}
\label{fig:OS}
\end{center}
\end{figure}

When $Ra$ is a few times larger than the critical value, convective motions become {quasi-periodic}. Figure \ref{fig:OS} shows an example of the {relaxation oscillation} regime at $E=10^{-5}$ and $Ra=1000\approx 4.8 Ra_c$. We can see that the time evolution of the energy exhibits a quasi-periodic manner (\ref{fig:OS}a). The growth of the onset mode (\ref{fig:OS}b) leads to subsequent instabilities, resulting in more vigorous and complex convection (\ref{fig:OS}c). A movie of the axial vorticity for this case is {provided} (see Movie 1). As the non-zonal energy approaches the maxima, there is a burst of the zonal energy, followed by a relaminarisation process (decreasing of both the zonal and non-zonal energy).  During these cycles,  the non-zonal energy can vary over several orders of magnitude between peaks and troughs.  Similar {relaxation oscillations} have been observed in numerical simulations of convection in spherical shells, {and are also referred to as convective bursts \citep{Grote2001,Busse2002,Christensen2002,Heimpel2012}}. The relaxation oscillations can be attributed to the competition between the nonlinear effect and the viscous effect \citep{Christensen2002}. The oscillatory regime exists only at intermediate Rayleigh numbers (filled circles in figure \ref{fig:RegimDiagram}). 

\begin{figure}
\begin{center}
\includegraphics[width=0.7 \textwidth]{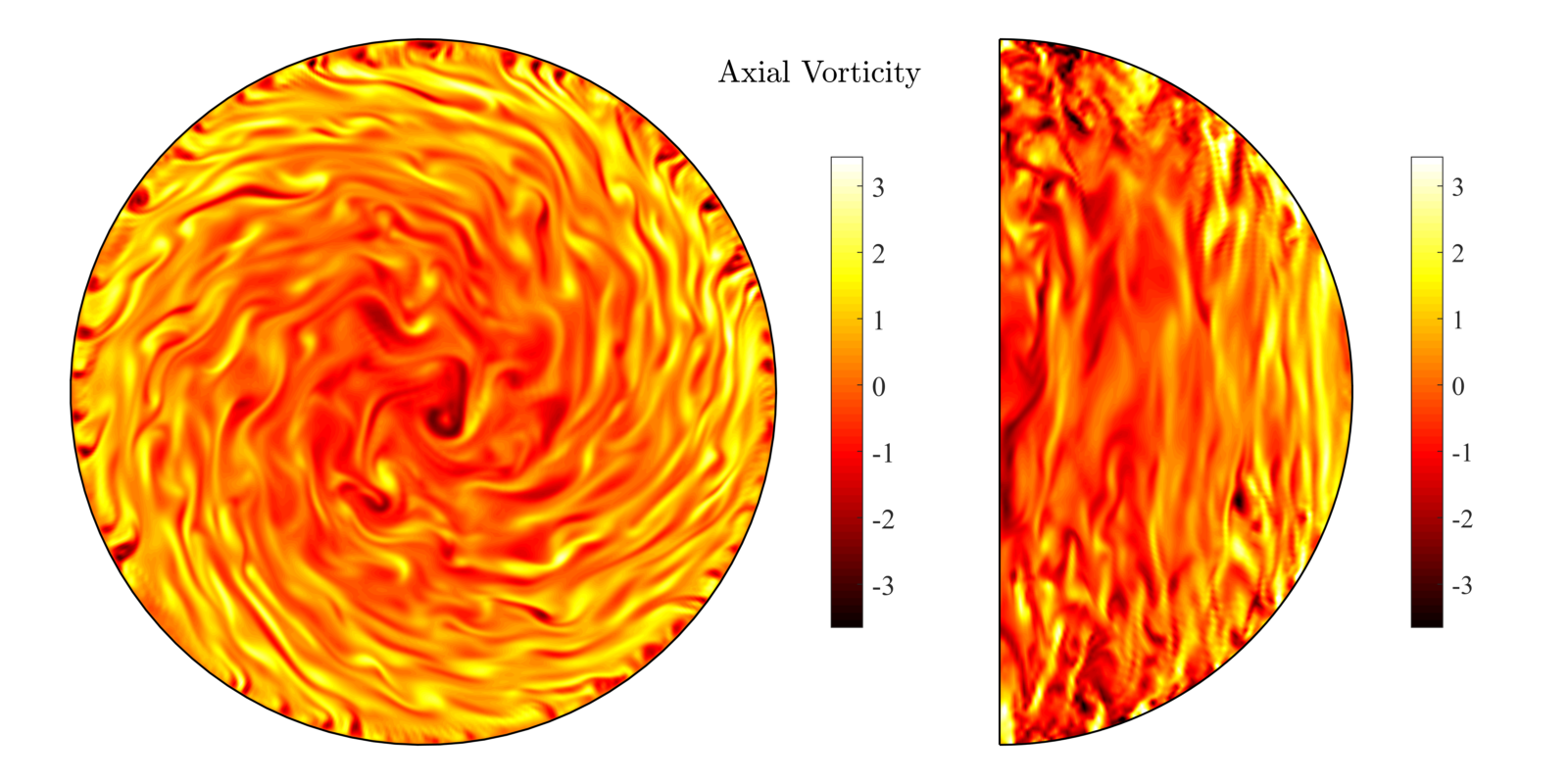}
\includegraphics[width=0.25 \textwidth]{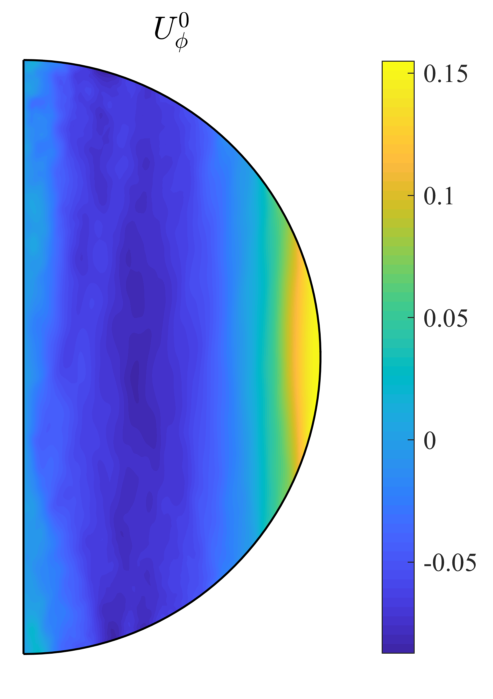} \\
(a) \hspace*{3cm} (b)\hspace*{3cm} (c) \\
\includegraphics[width=0.9 \textwidth]{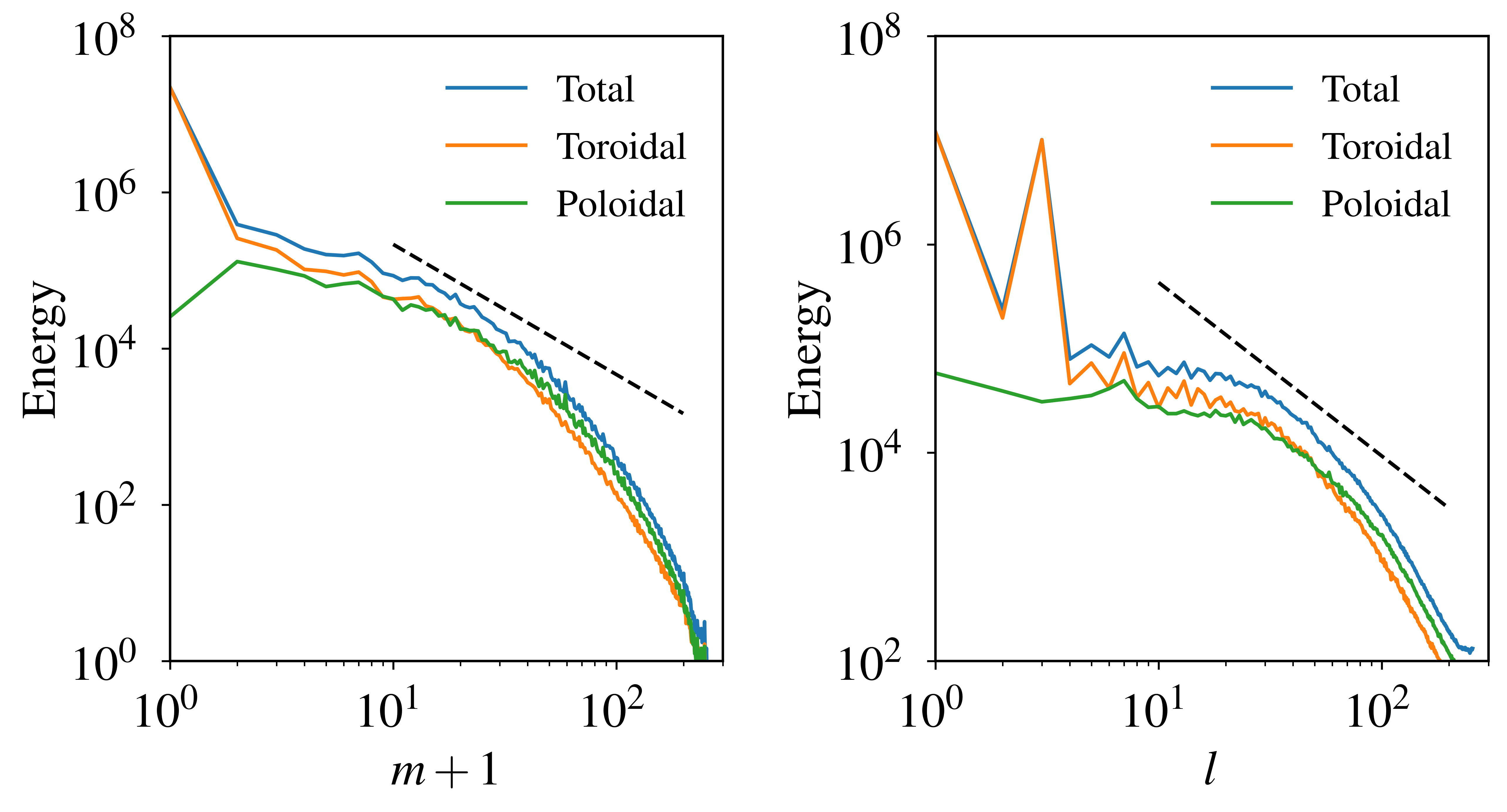} \\
(d) \hspace{5cm}(e) 
\caption{An example of the GT regime  at $E=10^{-5}$ and $Ra=10^5$. (a) Axial vorticity in the equatorial plane; (b) Axial vorticity in the meridional plane. {(c) Zonal flow in the meridional plane.} (d-e) Energy spectra as a function the azimuthal wavenumber $m$ and spherical harmonic degree $l$. Black dashed lines represent the $-5/3$ power law purely for reference. }
\label{fig:Chaotic}
\end{center}
\end{figure}

As we increase $Ra$,  time evolution of the kinetic energy becomes less periodic and eventually becomes irregular fluctuations, but the flow structures are still elongated along the rotation axis due to the constraint of rotation. { We refer to this regime as the GT regime}. The transition from {the relaxation oscillation to the GT regime} seems to be controlled by the convective Rossby number, i.e. $Ro_c\approx 0.2$ (blue dashed line in figure \ref{fig:RegimDiagram}). Figure \ref{fig:Chaotic} shows a typical case of the GT regime at $E=10^{-5}$ and $Ra=10^5$. In this regime, convective motions are charaterised by space-filling and sustained turbulence (\ref{fig:Chaotic}a-b). The axial vorticity in the meridional plane (\ref{fig:Chaotic}b) shows elongated structures along the rotation axis due to the constraint of rotation in this regime. This is a typical flow regime in rotating convection when $Ra$ is supercritical, which has been observed in a wide range of numerical simulations and laboratory experiments \citep{Aurnou2015}. 

Apart from small-scale turbulence, geostrophic zonal flows (figure \ref{fig:Chaotic}(c)) are developed in this case, with a strong prograde jet near the equator and mainly retrograde flows in the bulk. The significance of the zonal flow is clearly evident in the energy spectra (figure \ref{fig:Chaotic} (d-e)) as a function of the azimuthal wavenumber $m$  and spherical harmonic degree $l$. The kinetic energy is dominated by the $m=0$ toroidal component, which is basically the zonal flow. In the $l$ spectra, $l=0$ and $l=2$ toroidal components contribute the most kinetic energy in the flow.

\begin{figure}
\begin{center}
\includegraphics[width=0.7 \textwidth]{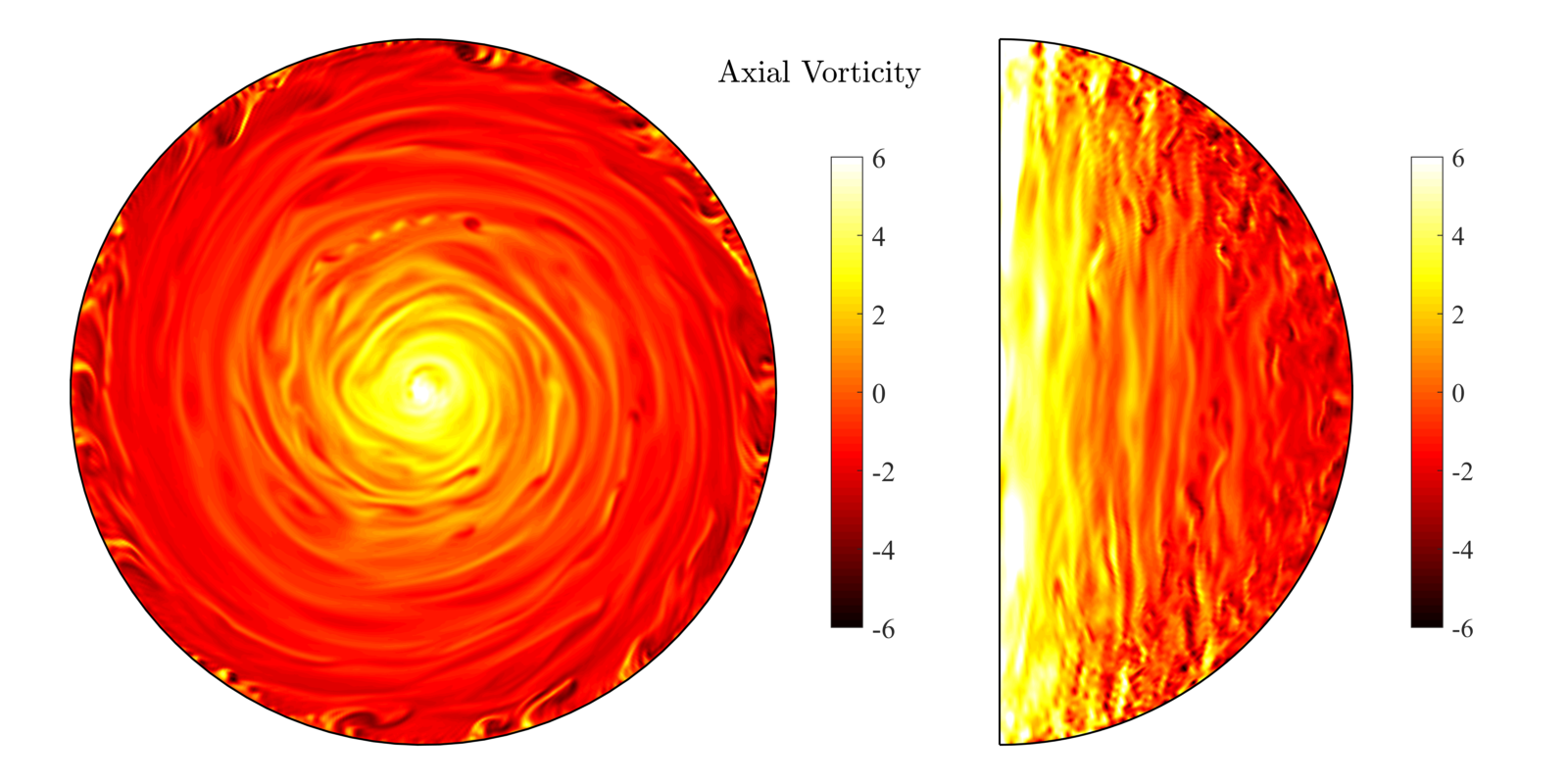}
\includegraphics[width=0.25 \textwidth]{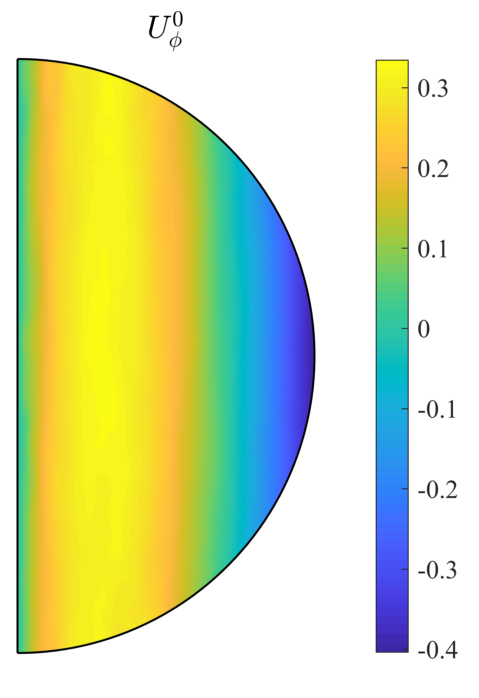} \\
(a) \hspace*{3cm} (b)\hspace*{3cm} (c) \\
\includegraphics[width=0.95 \textwidth]{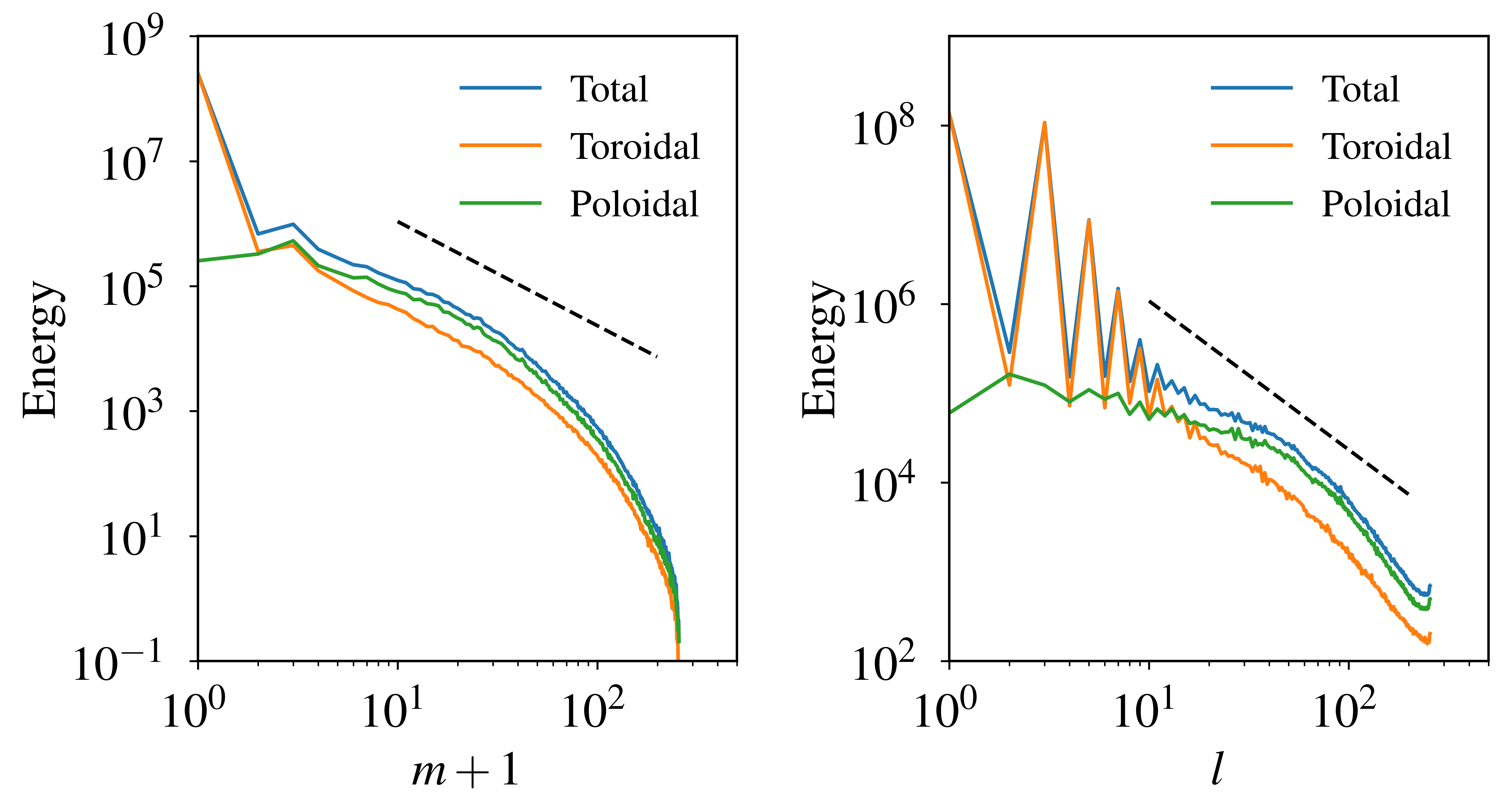} \\
(d)\hspace{5cm}(e)
\caption{An example of the LSV regime  at $E=10^{-5}$ and $Ra=3\times10^5$. (a) Axial vorticity in the equatorial plane; (b) Axial vorticity in the meridional plane. {(c) Zonal flow in the meridional plane.} (d-e) Energy spectra as a function the azimuthal wavenumber $m$ and spherical harmonic degree $l$. Black dashed lines represent the $-5/3$ power law purely for reference. }
\label{fig:LSV}
\end{center}
\end{figure}
Further increasing $Ra$ leads to a completely different flow regime (red triangles in figure \ref{fig:RegimDiagram}), which is characterised by the formation of LSV. The transition from the GT regime to the LSV regime is also determined by the convective Rossby number, i.e. $Ro_c\approx1.5$ (the red dashed line in figure \ref{fig:RegimDiagram}).  Figure \ref{fig:LSV} shows an example of the LSV  at $E=10^{-5}$ and $Ra=3\times10^5$.  We can see a notable cyclonic vortex located at the centre in the equatorial plane (figure \ref{fig:LSV}(a)) and extended along the rotation axis (figure \ref{fig:LSV}(b)). The formation process of the LSV is further discussed in section \ref{sec:LSV}. The vortex is accompanied by a strong geostrophic zonal flow (figure \ref{fig:LSV}(c)). However, the direction of the zonal flow is reversed compared to the previous regime. The zonal flow is retrograde near the equator and prograde in the inner region. The kinetic energy is dominated by the $m=0$ toroidal component, which corresponds to the columnar vortex and the zonal flow. The sawtooth-shaped curves in the $l$spectra at large $l$ reflect the equatorially symmetric nature of the LSV and the zonal flow.  

\begin{figure}
\begin{center}
\includegraphics[width=0.7 \textwidth]{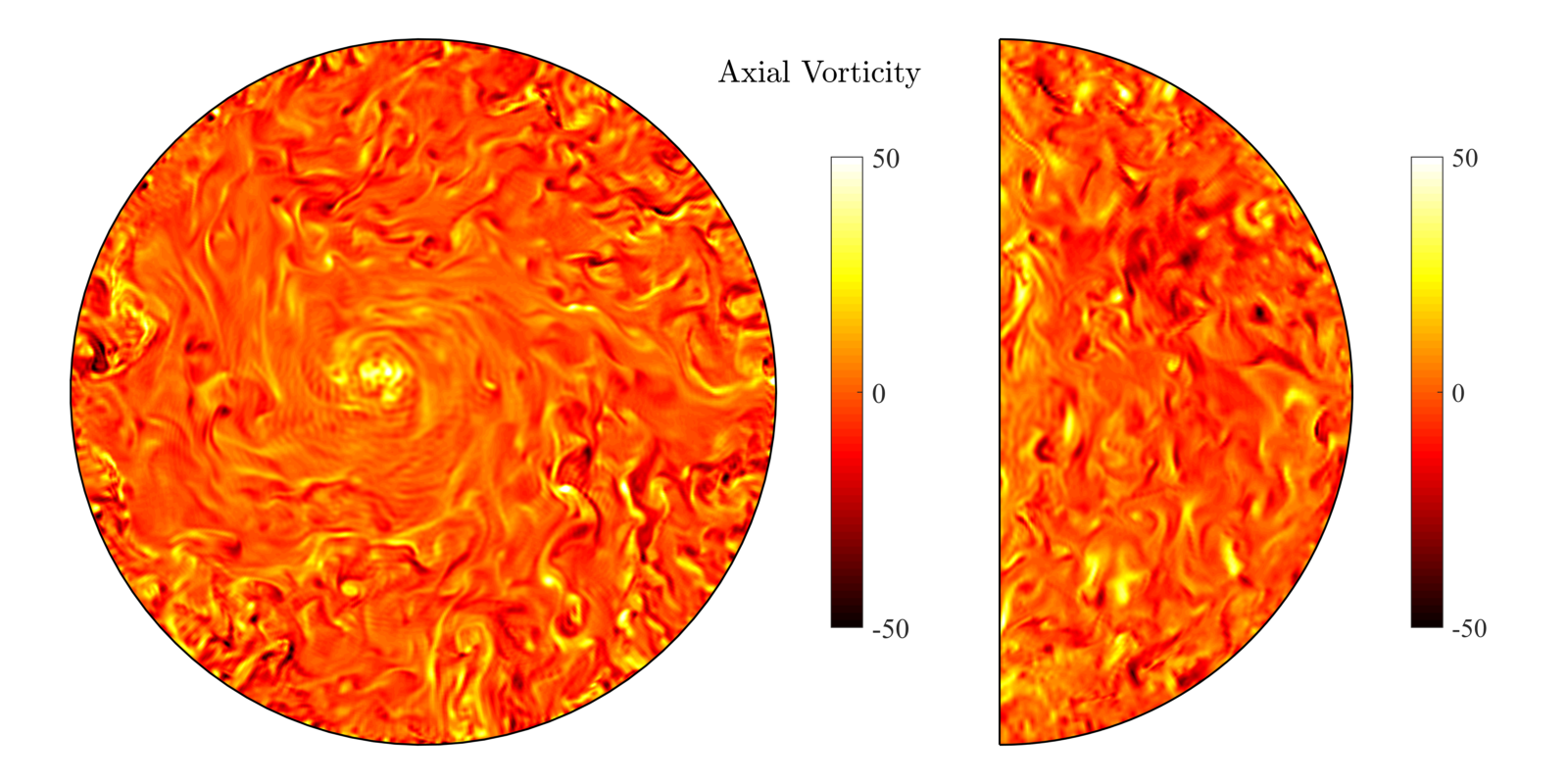}
\includegraphics[width=0.25 \textwidth]{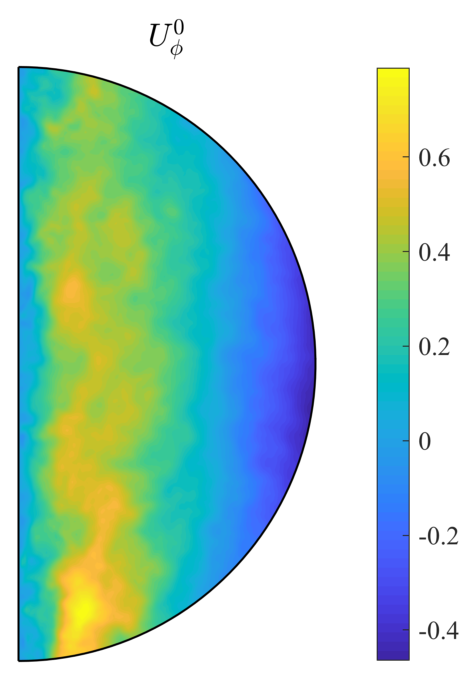} \\
(a) \hspace*{3cm} (b)\hspace*{3cm} (c) \\
\includegraphics[width=0.95 \textwidth]{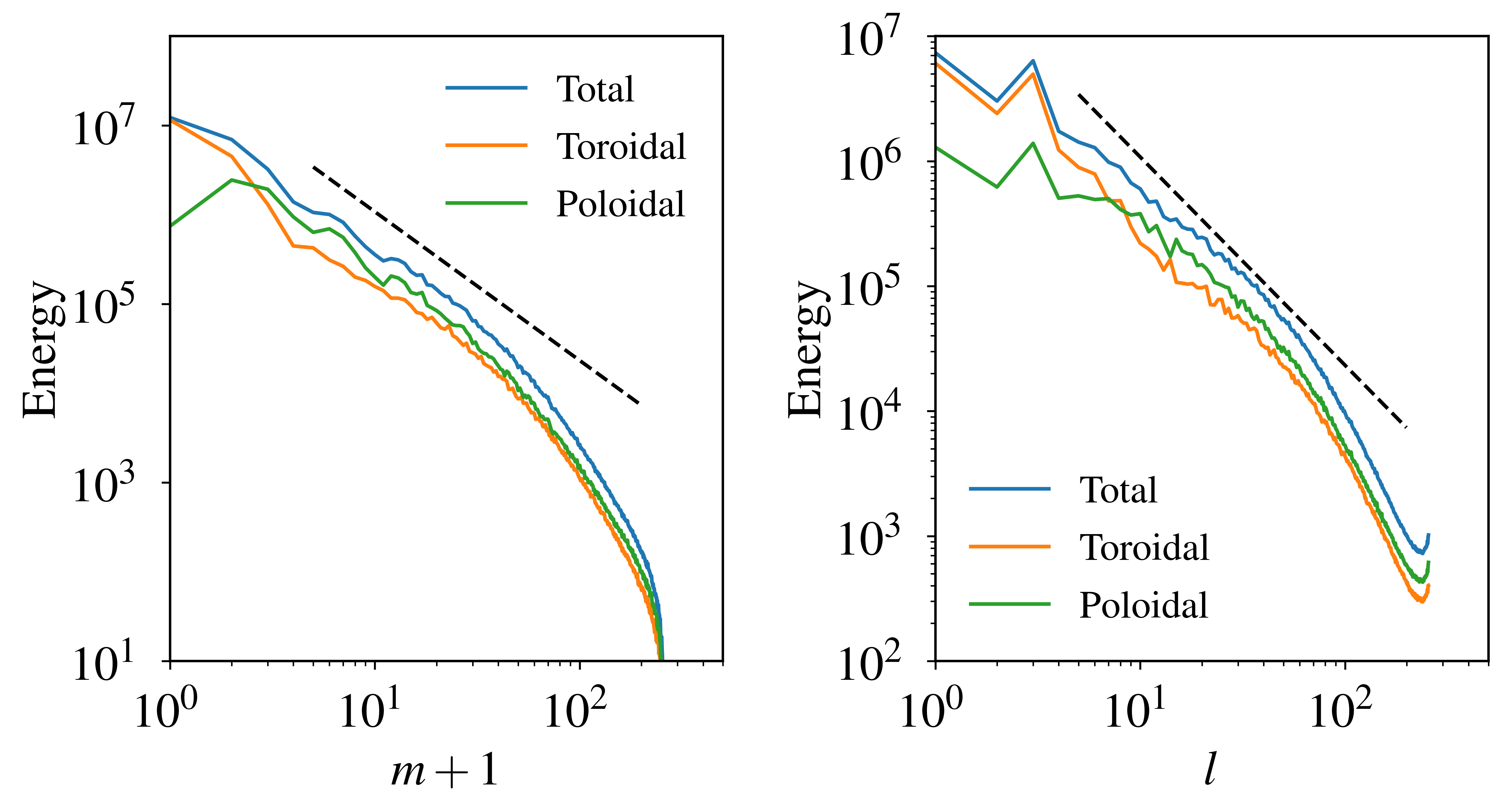} \\
(d)\hspace{5cm}(e)
\caption{ { An example of LSV breakdown} at $E=6.0\times 10^{-5}$ and $Ra=2.0\times10^6$. (a) Axial vorticity in the equatorial plane; (b) Axial vorticity in the meridional plane. {(c) Zonal flow in the meridional plane.} (d-e) Energy spectra as a function the azimuthal wavenumber $m$ and spherical harmonic degree $l$. Black dashed lines represent the $-5/3$ power law.}
\label{fig:LargeRa}
\end{center}
\end{figure}

At sufficiently large $Ra$, large-scale coherent structures break down and convection tends to behaviour similar to that  in a non-rotating regime. Figure \ref{fig:LargeRa} shows an example in this regime at $E=6.0\times 10^{-5}$ and $Ra=2.0\times 10^{6}\approx 1.6\times 10^4 Ra_c$ (corresponding to the open triangle in figure \ref{fig:RegimDiagram}). Unlike previous cases, axial vorticity in the equatorial and  meridional planes does not exhibit obvious anisotropy figure \ref{fig:LargeRa}(a-b). The zonal flow becomes much less geostrophic compared with previous cases (figure \ref{fig:LargeRa}(c)). Energy spectra in both $m$ and $l$ are in agreement with the Kolmogorov scaling  of $k^{-5/3}$ over a wide range of  wavenumbers (figure \ref{fig:LargeRa}(d-e)). This non-rotating regime exists at very large $Ra$ and it is computationally demanding to simulate this regime at lower $E$. As our focus in this study is the LSV, we do not explore   this regime further. 
\subsection{Formation of the LSV}
\label{sec:LSV}
\begin{figure}
\begin{center}
\includegraphics[width=0.9 \textwidth]{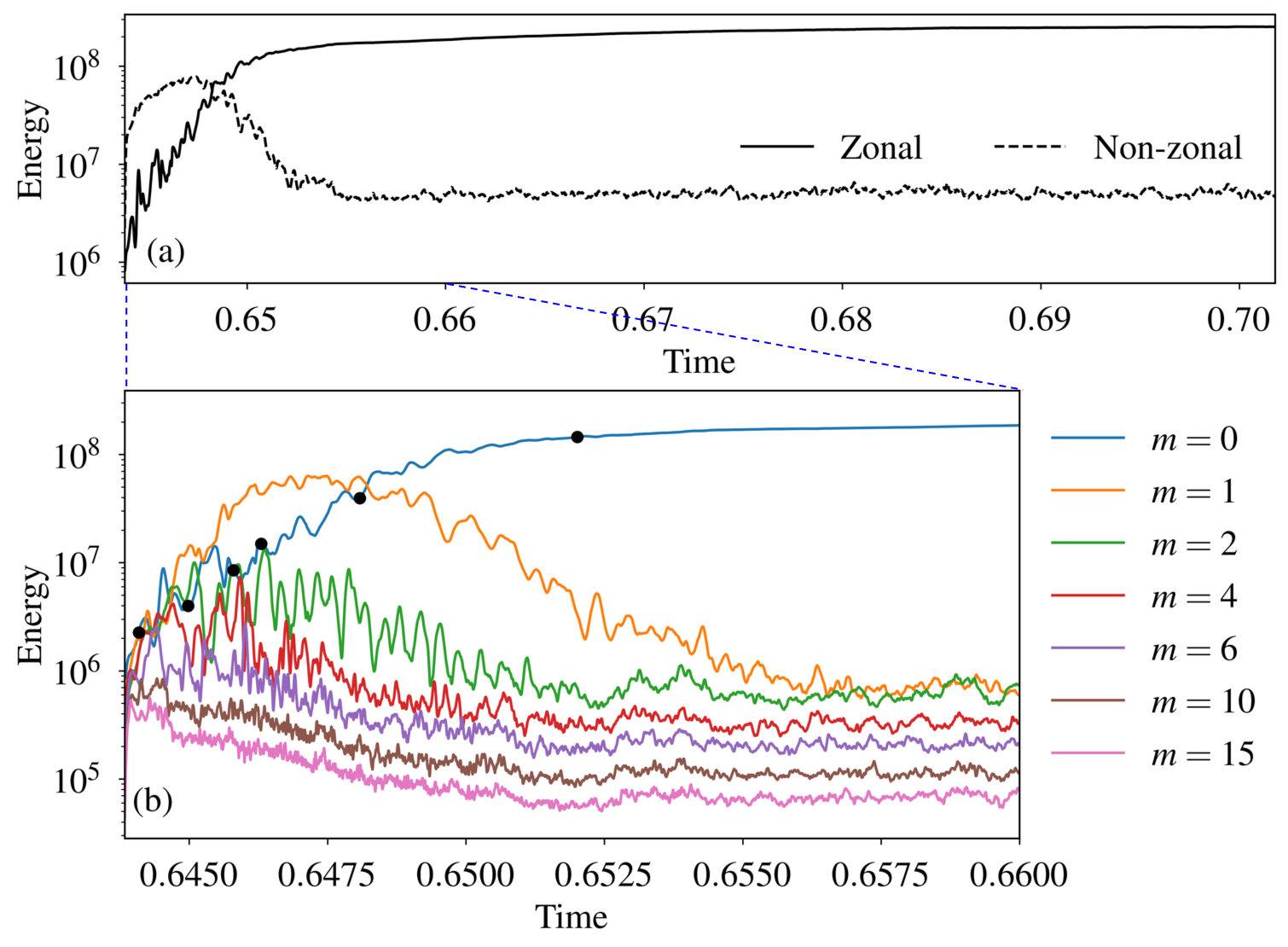}
\end{center}
\caption{Time series of energy at $E=10^{-5}$ and $Ra=3.0\times 10^{5}$. (a) Zonal and non-zonal energy. (b) Energy contained in several different azimuthal wavenumbers $m$ during the formation stage of the LSV. Time is in units of the viscous time scale.}
\label{fig:KEE1e-5Ra3e5Pr1}
\end{figure}

We now turn to analyse the formation process of the LSV. Figure \ref{fig:KEE1e-5Ra3e5Pr1} shows the time evolution of the zonal energy and the non-zonal energy (a), as well as the kinetic energy carried by several different $m$ during the growth phase of the LSV (b) at $E=10^{-5}$ and $Ra=3.0\times 10^{5}$ (corresponding to the case in figure \ref{fig:LSV}). The simulation started from a saturated state at lower $Ra$ in the GT regime. We can see that the zonal energy overtakes the non-zonal energy after the transient stage and then the non-zonal energy drops. The simulation saturates to a state of which the zonal energy is dominant over the non-zonal energy. 

The energy transfer from a small-scale flows to a large-scale flow is probably best illustrated in figure \ref{fig:KEE1e-5Ra3e5Pr1} (b). While the energy in the $m=0$ component keeps growing and becomes dominant, the energy in other components initially grows but subsequently drops. We note that the larger the azimuthal wavenumber $m$, the earlier the energy drop, suggesting a successive energy transfer from larger $m$ to smaller $m$. Similar upscale energy transfer has been reported in   rotating Rayleigh-B\'enard convection, leading to box-sized vortices \citep{Guervilly2014,Rubio2014}. In the full sphere we consider here, the upscale energy transfer leads to a large cyclonic vortex located at the centre and strong zonal flow 
(figure \ref{fig:LSV}). 

\begin{figure}
\begin{center}
\includegraphics[width=0.48\textwidth]{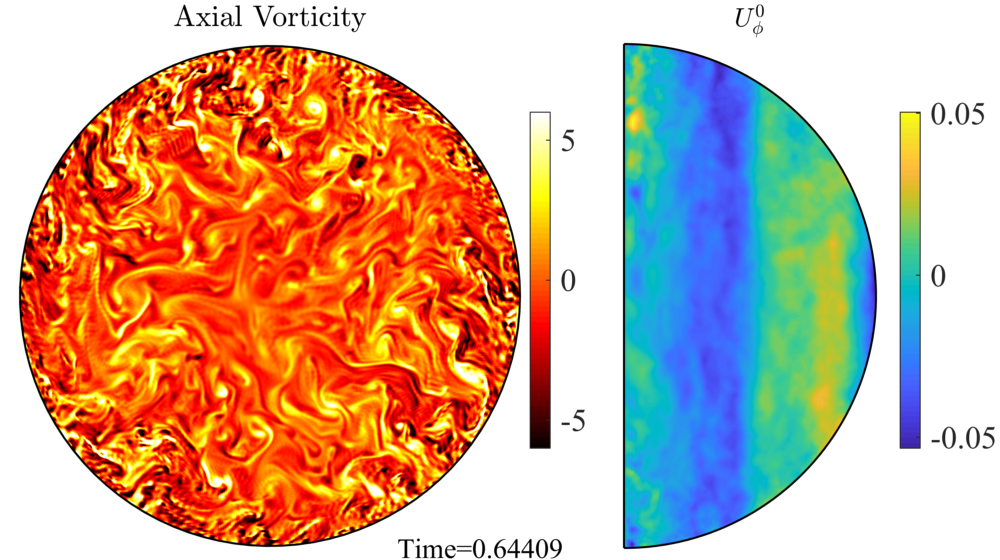}
\includegraphics[width=0.48\textwidth]{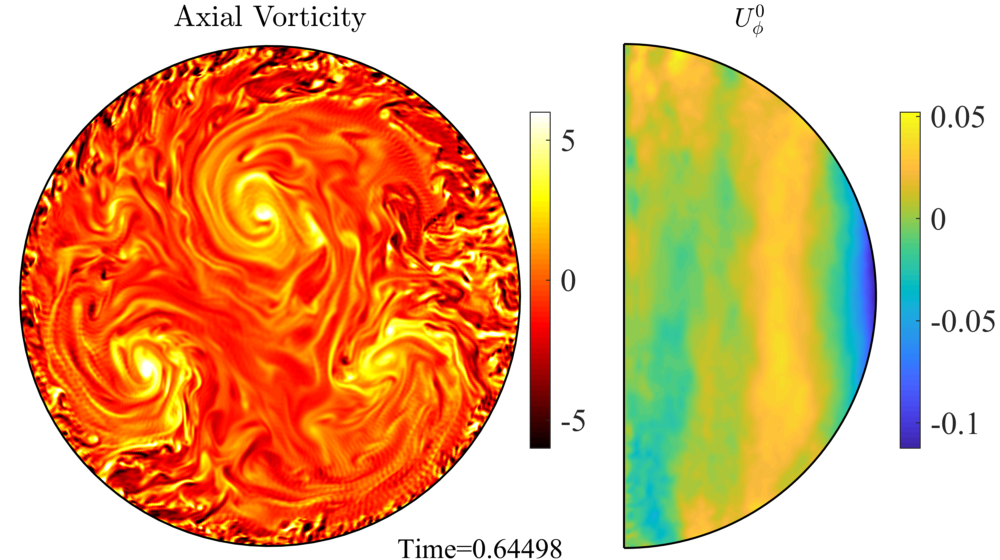}
\\
(a)\hspace{5cm}(b) \\
\includegraphics[width=0.48\textwidth]{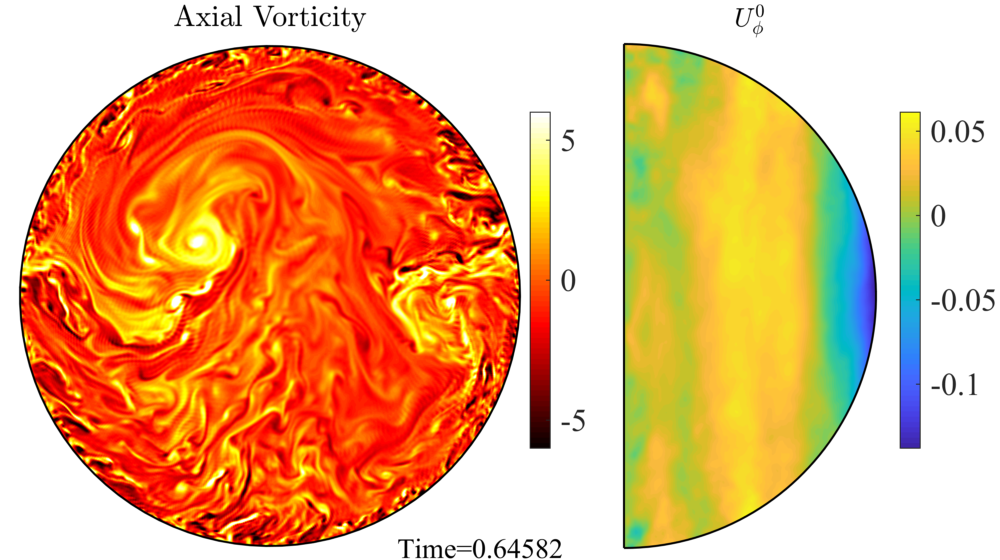}
\includegraphics[width=0.48\textwidth]{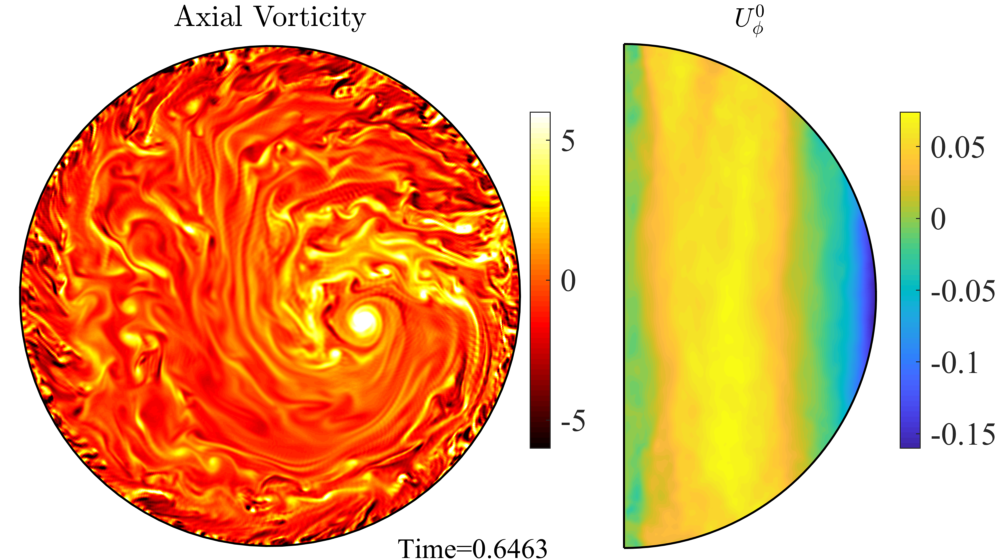}
\\
(c)\hspace{5cm}(d) \\
\includegraphics[width=0.48\textwidth]{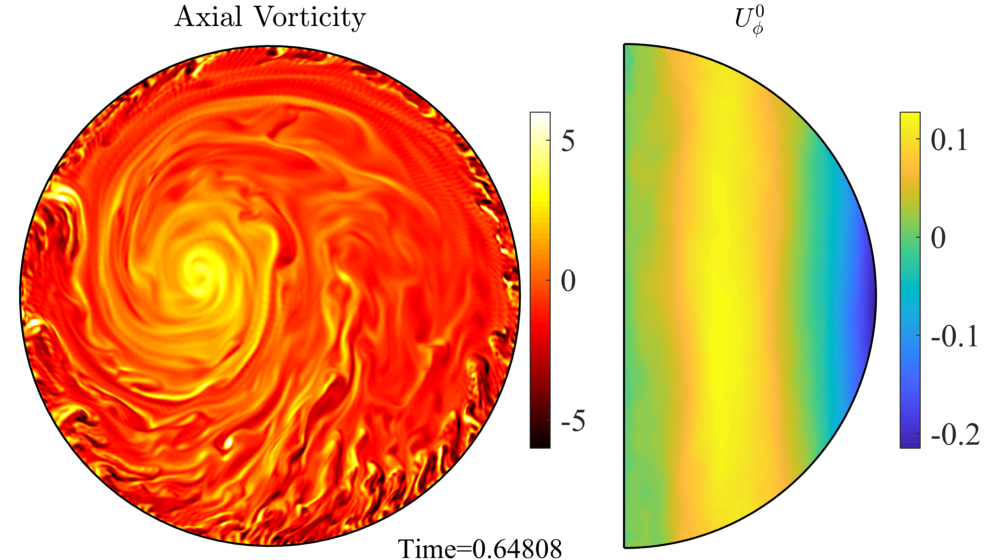}
\includegraphics[width=0.48\textwidth]{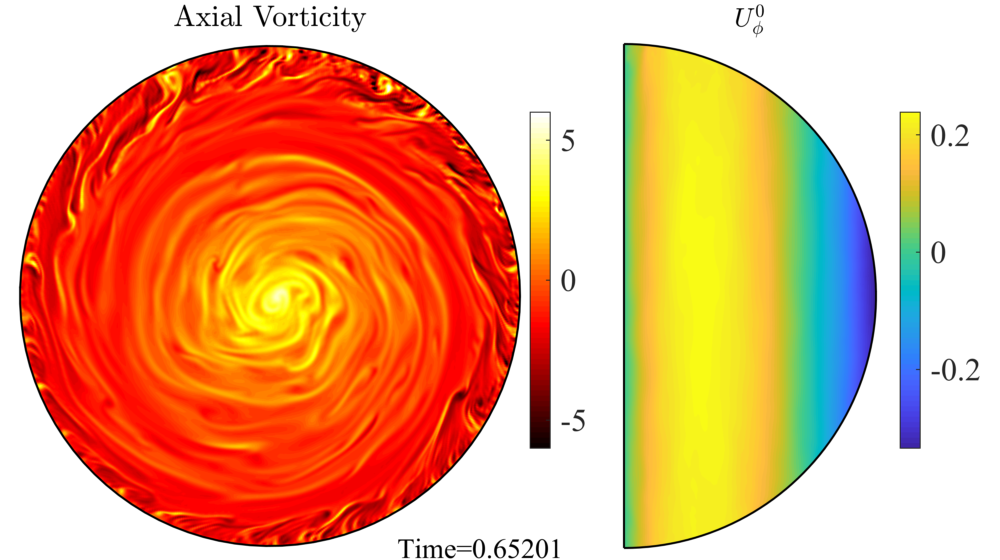}
\\
(e)\hspace{5cm}(f) 
\end{center}
\caption{Snapshots (corresponding to the black dots in figure \ref{fig:KEE1e-5Ra3e5Pr1}(b)) of the axial vorticity in the equatorial plane and zonal flow in the meridional plane at $E=10^{-5}$ and $Ra=3.0\times 10^{5}$. (A movie of this time series is provided; see \href{https://faculty.sustech.edu.cn/wp-content/uploads/2020/12/2020122222193447.mp4}{Movie 2}).}
\label{fig:VorZE1e-5Ra3e5Pr1}
\end{figure}

Figure \ref{fig:VorZE1e-5Ra3e5Pr1} shows a sequence of snapshots (corresponding to the black dots in figure \ref{fig:KEE1e-5Ra3e5Pr1}(b)) of the axial vorticity and the zonal flow during the formation stage of the LSV at $E=10^{-5}$ and $Ra=3.0\times 10^{5}$. We clearly see the process of vortex merging from small scales (figure \ref{fig:VorZE1e-5Ra3e5Pr1}(a)) to a few prominent vortices (figure \ref{fig:VorZE1e-5Ra3e5Pr1} (b-c)), which further merge into one dominant cyclonic vortex (figure \ref{fig:VorZE1e-5Ra3e5Pr1} (d)). The eye of the cyclone spirals inwards and eventually settles at the centre (figure \ref{fig:VorZE1e-5Ra3e5Pr1} (e-f)).  The vortex is elongated along the rotation axis as we have shown in figure \ref{fig:LSV} (b). As the vortices merge and migrate inwards, the prograde zonal flow also shifts inwards, suggesting an inwards transport of  angular momentum. We further discuss the mean zonal flow in section \ref{sec:zonal}.    

\begin{figure}
\begin{center}
\includegraphics[width=0.49\textwidth]{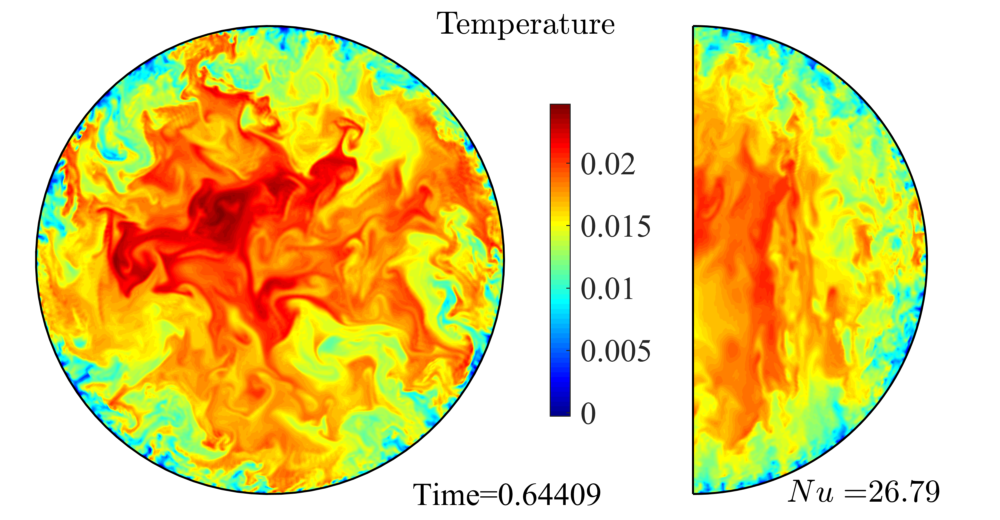}
\includegraphics[width=0.49\textwidth]{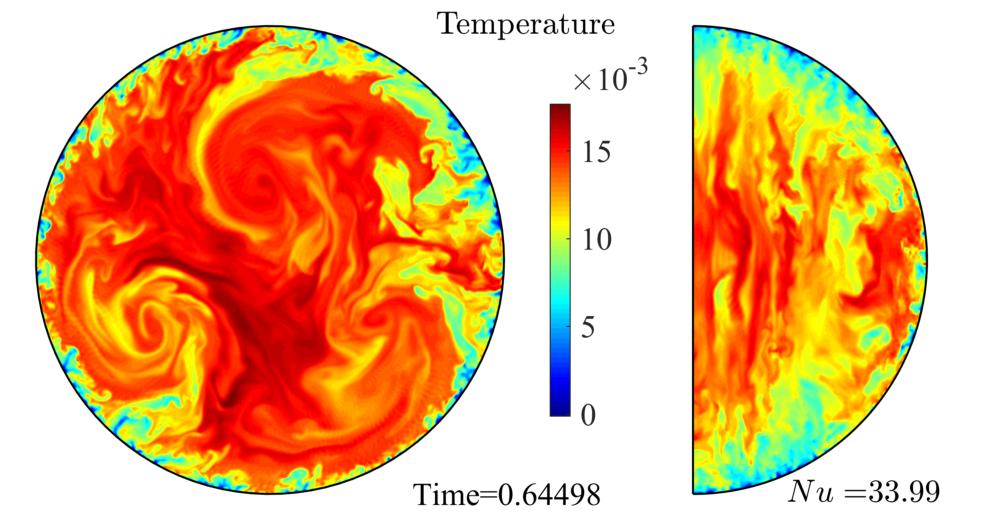}\\
(a)\hspace{5cm}(b) \\
\includegraphics[width=0.49\textwidth]{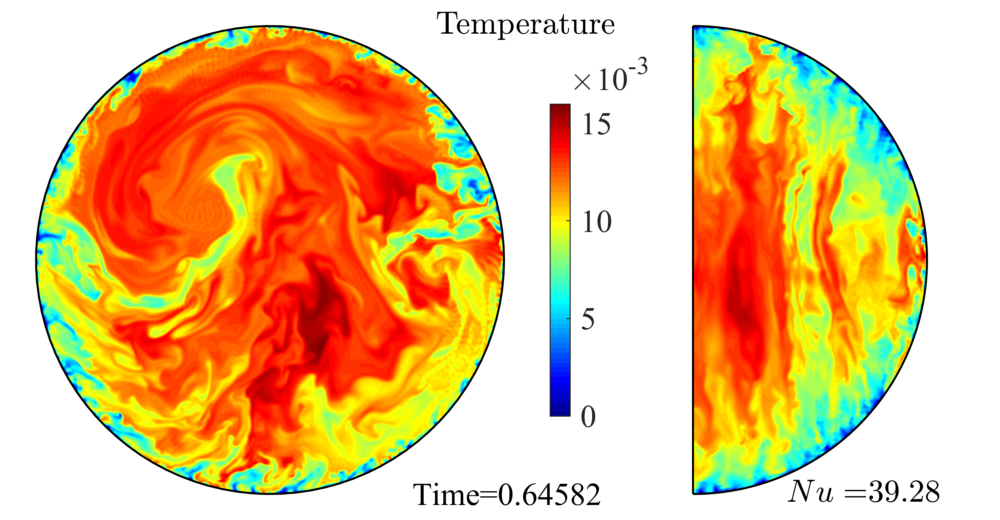}
\includegraphics[width=0.49\textwidth]{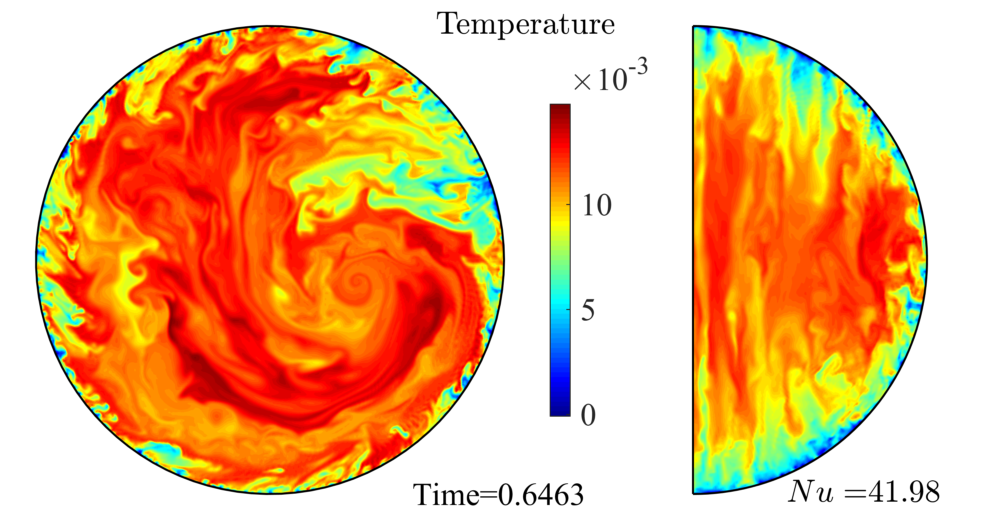} \\
(c)\hspace{5cm}(d) \\
\includegraphics[width=0.49\textwidth]{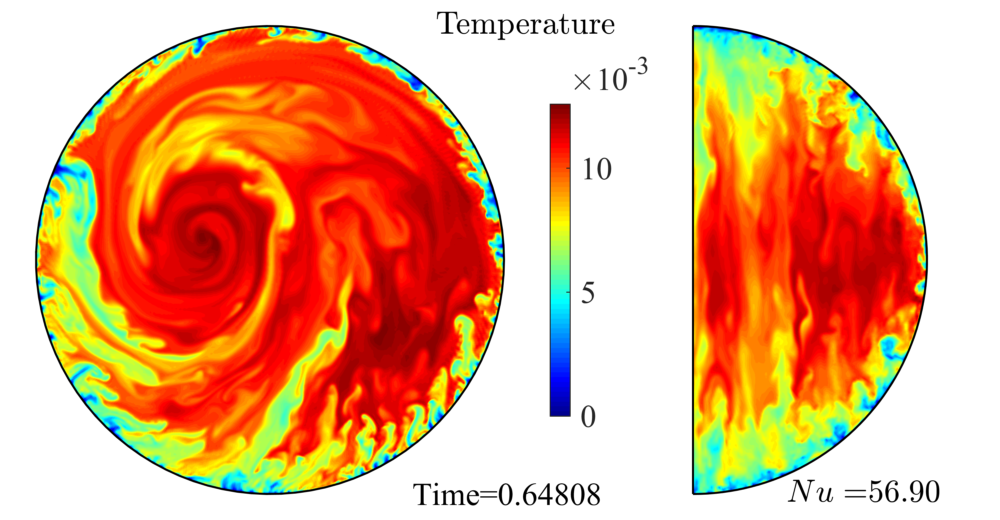}
\includegraphics[width=0.49\textwidth]{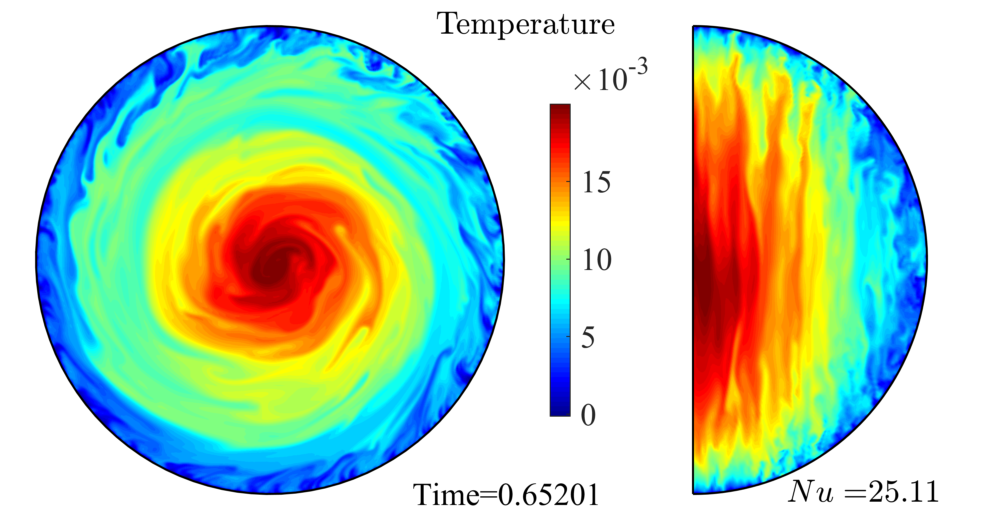} \\
(e)\hspace{5cm}(f)
\end{center}
\caption{As for figure \ref{fig:VorZE1e-5Ra3e5Pr1} but for the temperature in the equatorial plane and in the meridional plane. Instantaneous Nusselt number $Nu$ is also indicated.}
\label{fig:TempE1e-5Ra3e5Pr1}
\end{figure}

Figure \ref{fig:TempE1e-5Ra3e5Pr1} shows the same snapshots as in figure \ref{fig:VorZE1e-5Ra3e5Pr1} but for the temperature field. We also see the merging process from small-scale to  larger-scale structures of the temperature field, culminating with a relatively hot columnar aggregation around the rotation axis as the LSV is formed. The instantaneous  Nusselt number $Nu$ is  indicated at the bottom of each snapshot. We note that the formation of the LSV results in a significant drop of $Nu$, e.g. from $Nu=56.9$ (in figure \ref{fig:TempE1e-5Ra3e5Pr1}e) to $Nu=25.11$ (in figure \ref{fig:TempE1e-5Ra3e5Pr1}f). This is perhaps due to the fact that  a hot column of fluid is trapped around the rotation axis by the LSV, which reduces the amount of heat transported by the convection. We  further discuss effects of the LSV on the heat transport in section \ref{sec:Nu}.

\subsection{Mean zonal flows}\label{sec:zonal}
As we have shown, zonal flows are always developed once the convective motions set in. In this section, we compare the zonal flow in different regimes and investigate how the zonal flow varies depending on the control parameters. 
\begin{figure}
\begin{center}
\includegraphics[width=0.32\textwidth]{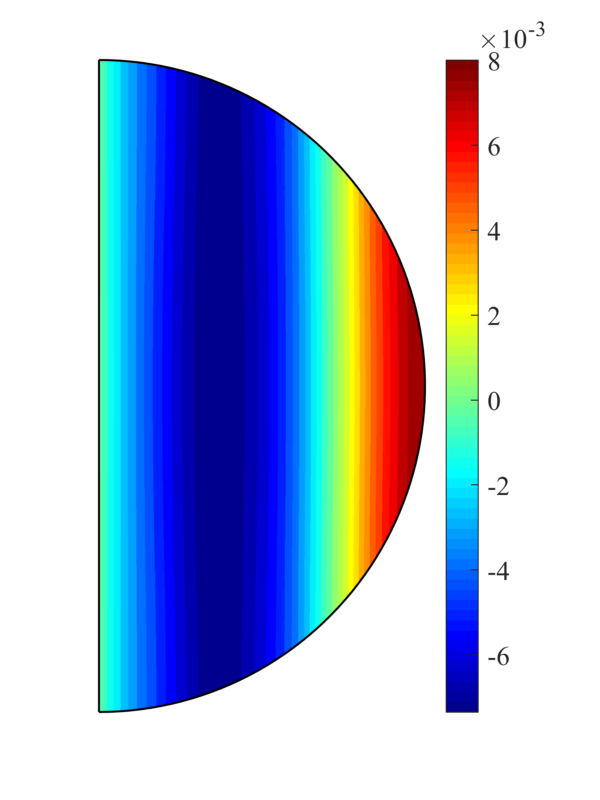}
\includegraphics[width=0.32\textwidth]{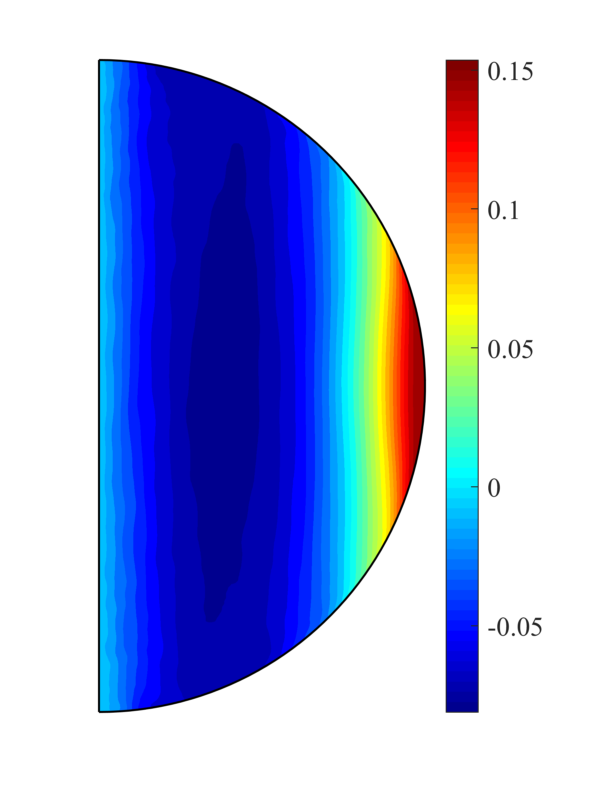}
\includegraphics[width=0.32\textwidth]{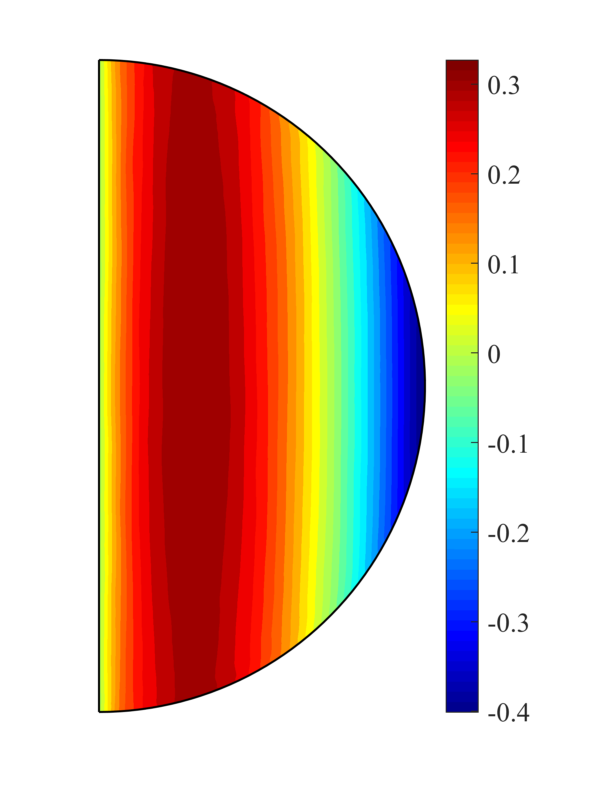} \\
(a)\hspace{3cm}(b)\hspace{3cm}(c)
\end{center}
\caption{Mean zonal flows in the meridional plane at different $Ra$ for fixed $E=10^{-5}$. (a) $Ra=1.0\times 10^{3}, ~Ro_c=0.1$, in the { relaxation oscillation} regime; (b) $Ra=1.0\times 10^{5},~ Ro_c=1.0$, in the GT regime; (c) $Ra=3.0\times 10^{5}, ~Ro_c=\sqrt{3}$, in the LSV regime.}
\label{fig:ZonalPr1}
\end{figure}

Figure \ref{fig:ZonalPr1} shows the mean zonal flow, i.e. time-averaged zonal flow $\overline{U_\phi^0}$, in the meridional plane at various $Ra$ but fixed $E=10^{-5}$. In all cases, the mean zonal flows are almost invariant along the rotation axis. However, the direction of the mean zonal flow reverses when the convective Rossby number is sufficiently large such that the LSV forms, i.e. $Ro_c \gtrsim 1.5$. Similar reversal of the zonal flow has been observed in thin shells for both Boussinesq convection \citep{Aurnou2007} and anelastic convection { \citep{Gastine2013}}. They also found that the reversal is controlled by the convective Rossby number $Ro_c$, or equivalently by $Ra^*$.

\begin{figure}
\begin{center}
\includegraphics[height=0.32 \textwidth]{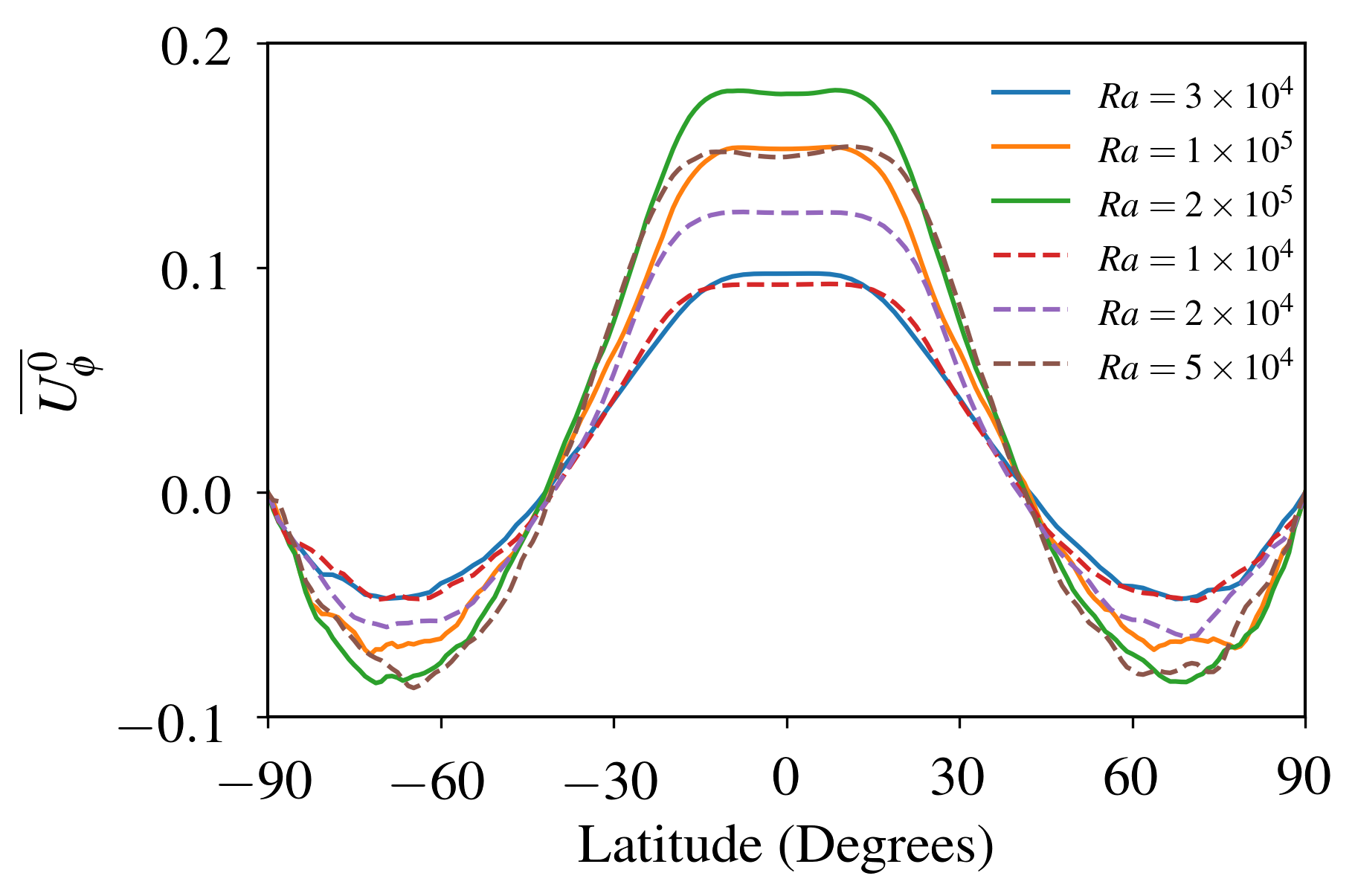}
\includegraphics[height=0.32 \textwidth]{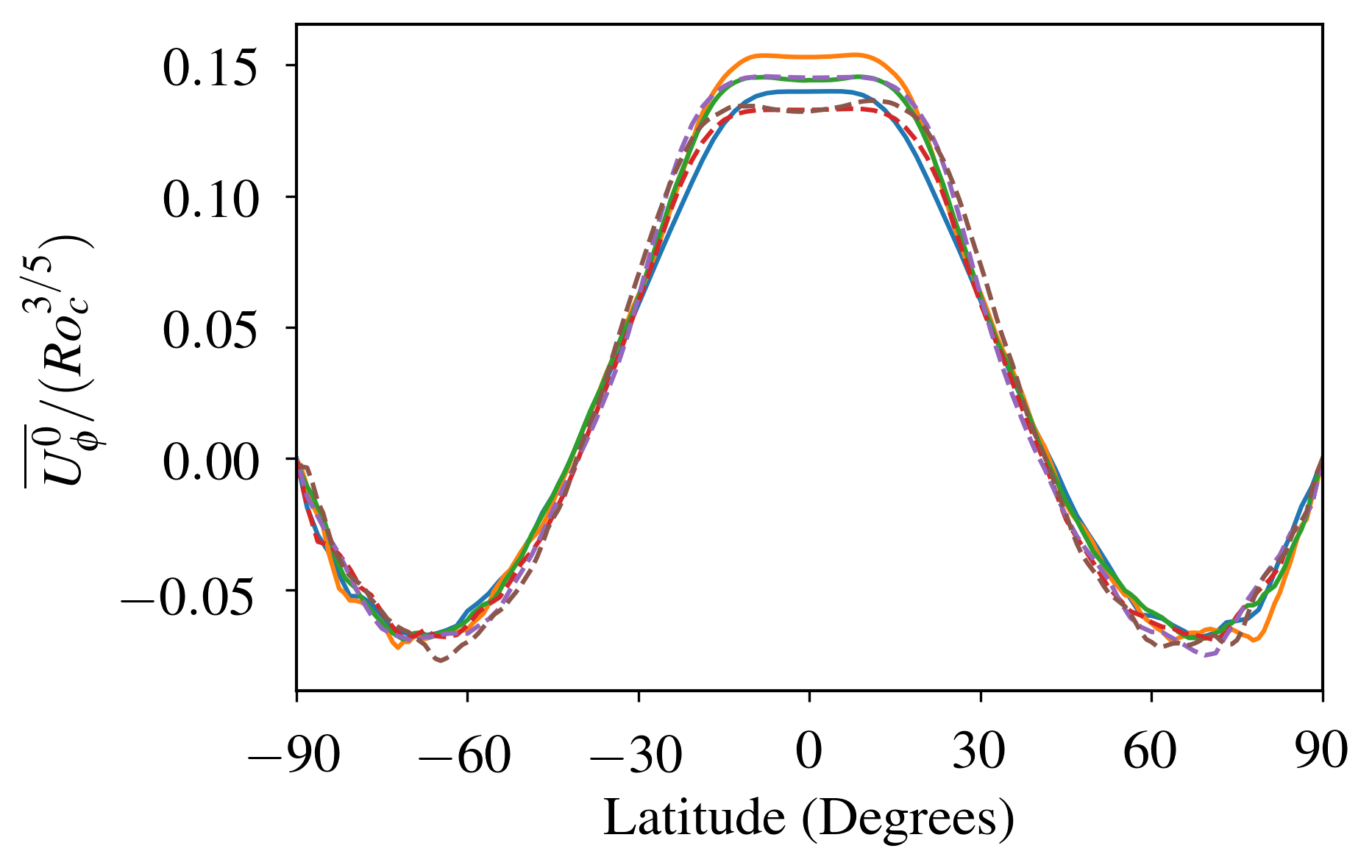}\\
(a) \hspace{6cm} (b)
\caption{(a) Mean zonal flow on the surface as a function of the latitude at various $E$ and $Ra$, but all cases in the GT regime. (b) As for (a) but the mean zonal flow is rescaled by $Ro_c^{3/5}$. Solid (dashed) lines correspond to cases at $E=10^{-5}$ ($E=3\times 10^{-5}$).}
\label{fig:ZonalProfilePr1}
\end{center}
\end{figure}
As the mean zonal flow is essentially geostrophic, we use the zonal velocity profile in the equatorial plane to represent the zonal flow. Figure \ref{fig:ZonalProfilePr1} (a) shows the zonal profiles at various $E$ and $Ra$ in the GT regime. These profiles collapse into an invariant profile when the zonal velocity is divided by $Ro_c^{3/5}$. {Note that the exponent $3/5$ is merely an empirical estimate and may be subject to small revisions.} Nevertheless, it is clear that the zonal flow is controlled by the the convective Rossby number in the GT regime.

\begin{figure}
\begin{center}
\includegraphics[height=0.33 \textwidth]{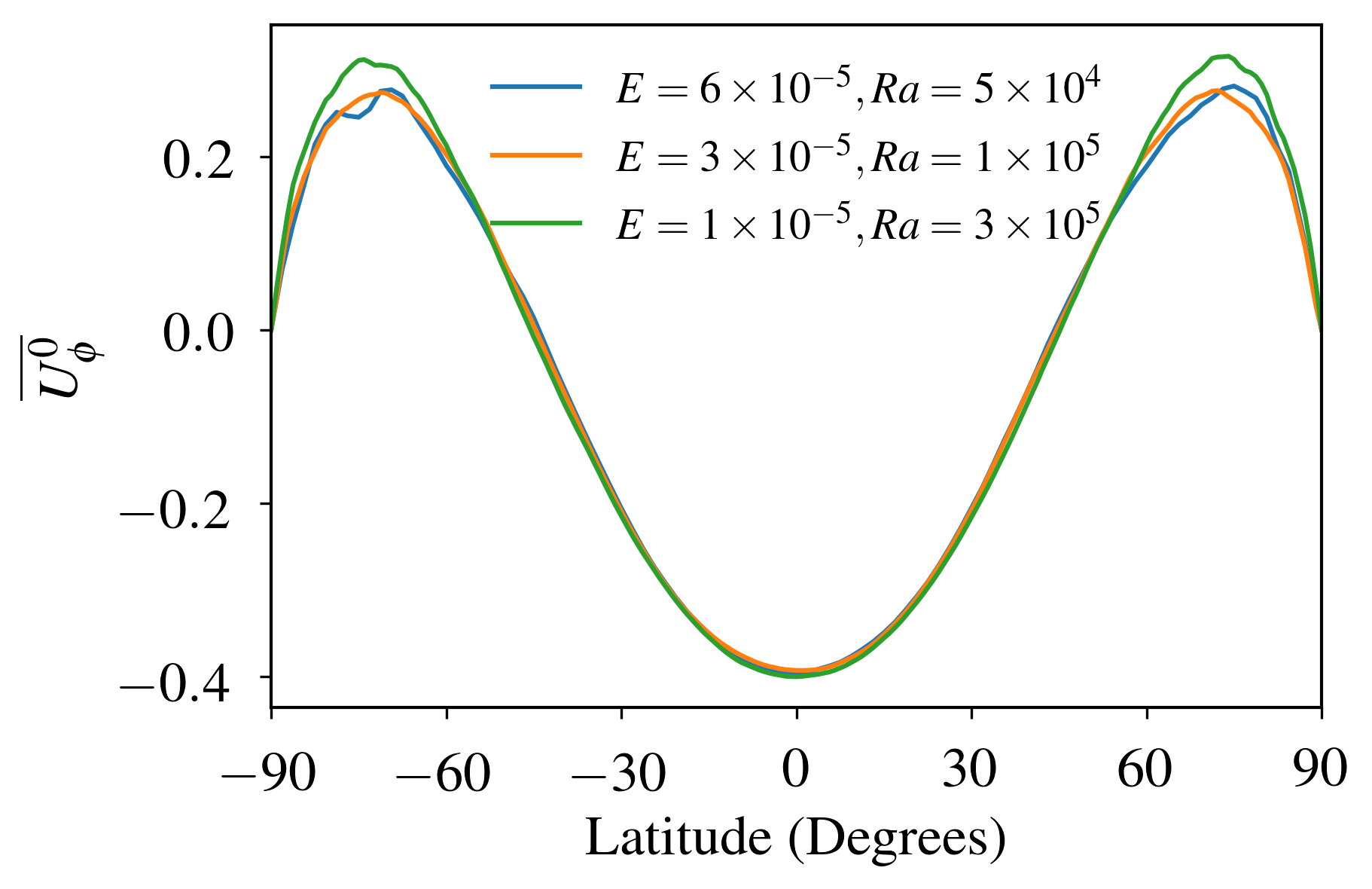}
\hfill
\includegraphics[height=0.33 \textwidth]{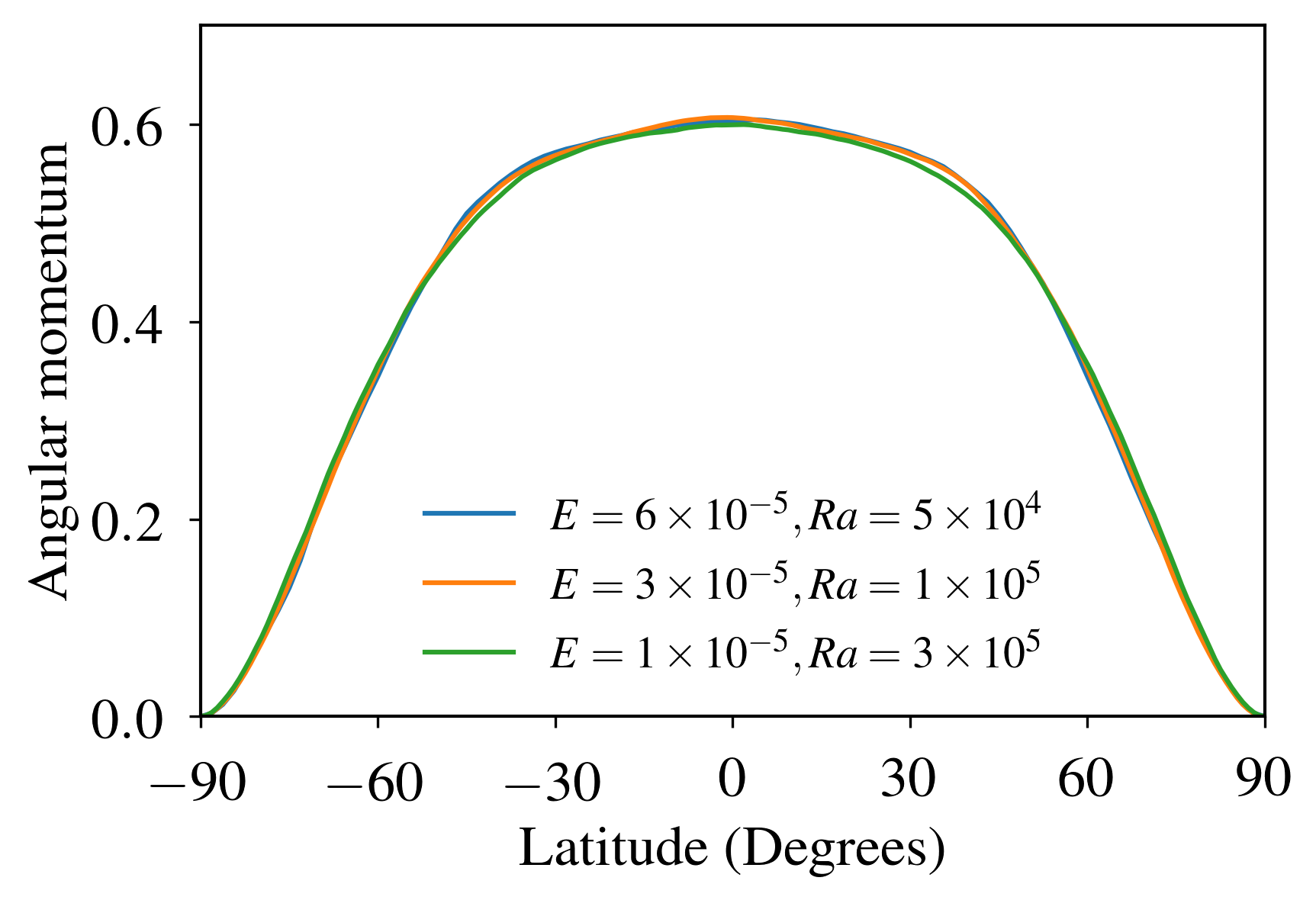} \\
(a) \hspace{6cm} (b)
\caption{(a) Mean zonal flow on the surface as a function of the latitude at various $E$ and $Ra$, but all cases in the LSV regime. As for (a) but showing the specific angular momentum in the inertial frame.}
\label{fig:ZonalProfilePr1LSV}
\end{center}
\end{figure}

\begin{figure}
\begin{center}
\includegraphics[width=0.48 \textwidth]{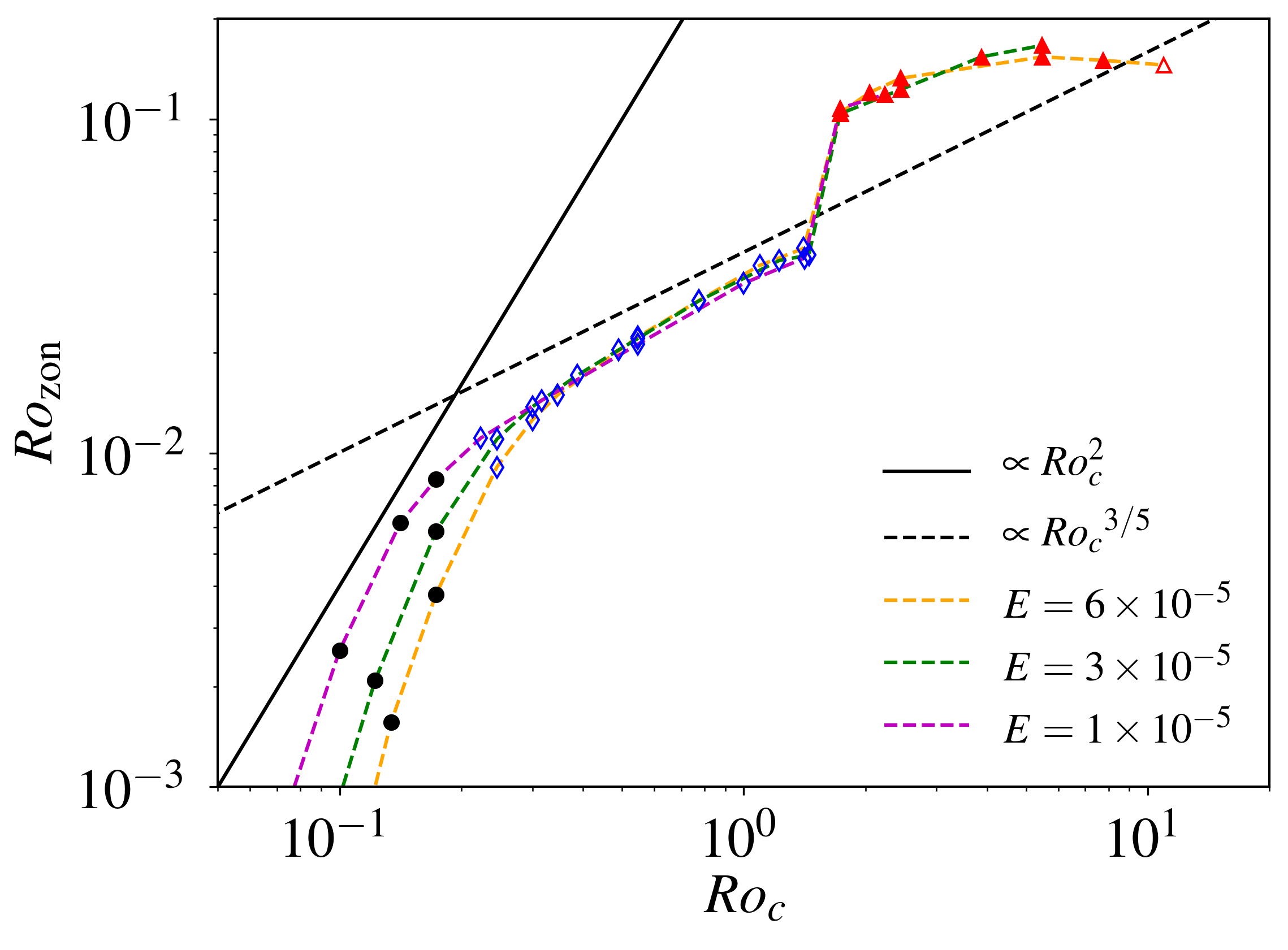}
\includegraphics[width=0.48 \textwidth]{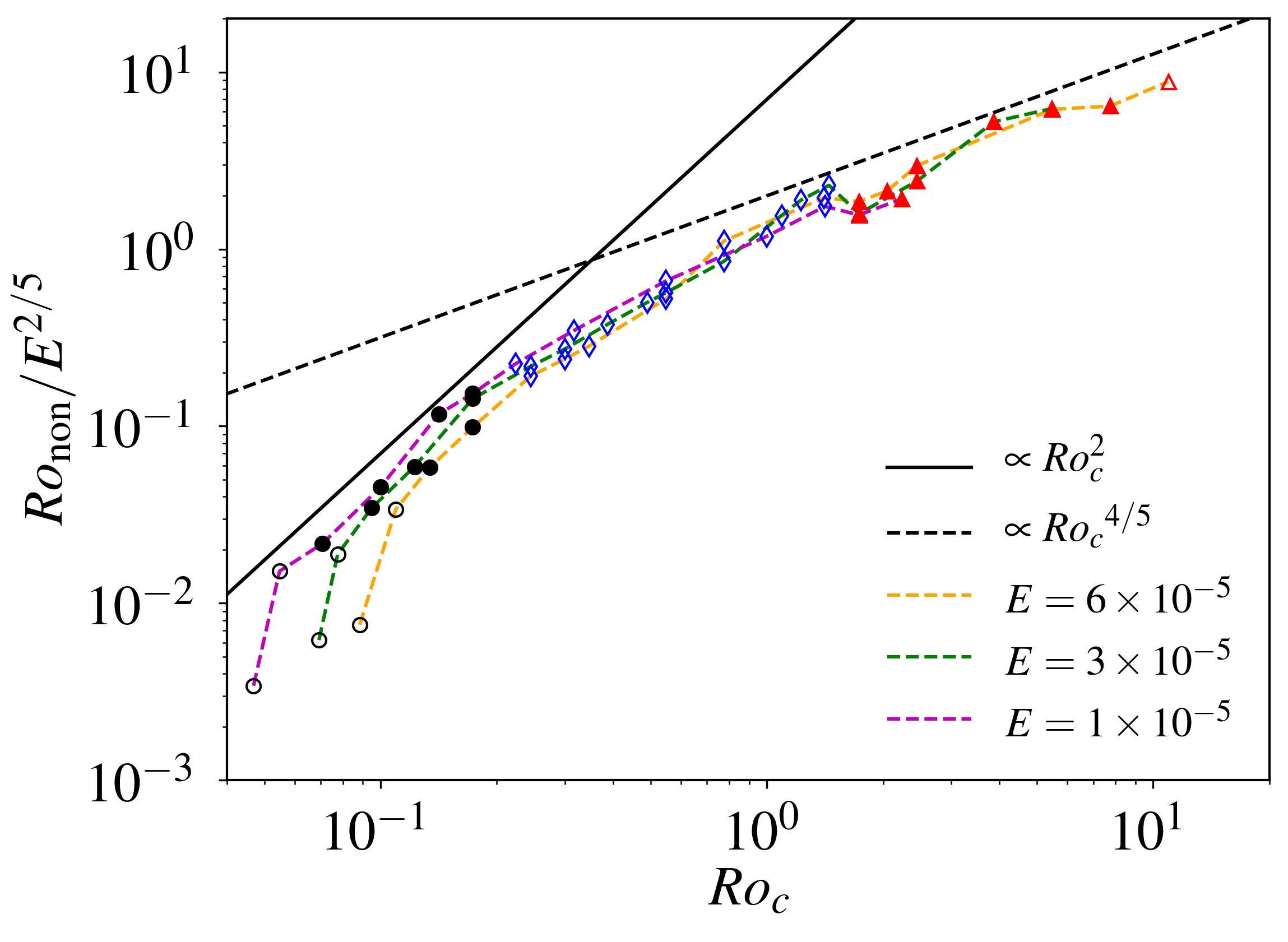} \\
(a)\hspace{6cm}(b)
\caption{Zonal Rossby number $Ro_{zon}$ (a) and {rescaled} non-zonal Rossby number $Ro_{non}/E^{2/5}$ (b) as a function of the convective Rossby number $Ro_c$.  Different symbols represent different flow regimes as in figure \ref{fig:RegimDiagram}. $Ro_{zon}$ and $Ro_{non}$ are time-averaged over several convective turnover time scales.}
\label{fig:Rossby}
\end{center}
\end{figure}

Figure \ref{fig:ZonalProfilePr1LSV}(a) shows the zonal profiles in the LSV regime for various $E$ and $Ra$, but the same $Ro_c=\sqrt{3}$.  
The zonal flow profile in the LSV regime (figure \ref{fig:ZonalProfilePr1LSV} (a)) is almost, but not exactly, the reverse of the profile in the GT regime (figure \ref{fig:ZonalProfilePr1} (b)). \citet{Aurnou2007} proposed that the vigorous convection at large $Ra$ homogenises the angular momentum in the inertial frame, leading to a retrograde equatorial jet and a prograde zonal {flow} inside in the rotating frame. Figure \ref{fig:ZonalProfilePr1LSV} (b) shows the specific angular momentum in the inertial frame, i.e. $M_I=s^2+s\overline{U_\phi^0}$ (non-dimensional). We see an inward transfer of the angular momentum compared to solid body rotation, but the angular momentum profiles are far from a homogeneous distribution, even in the outer region. One may expect a more homogeneous distribution of the angular momentum on further increasing $Ra$, but the zonal flow appears to saturate in the LSV regime as we can see from figure \ref{fig:Rossby} (a).     

We have been unable to 
discover a principle that predicts this profile. Of interest is the fact that the angular momentum in the LSV regime is clearly prograde for $s\lessapprox 0.72$ and retrograde for $s\gtrapprox 0.72$, with the sum being zero, of course, in the
rotating frame. But what determines this changeover point $s^\ast$? We merely remark that 
a fluid in solid body rotation contains half its angular momentum in $s\le 0.77$ and half exterior to this radius. Any connection between $s^\ast$ and this value may be fortuitous. 

The zonal Rossby number $Ro_{zon}$ increases with increasing $Ro_c$ with varying slopes depending on the flow regimes, and becomes saturated ($Ro_{zon}\sim 0.15$) in the LSV regime (figure \ref{fig:Rossby}(a)). The plot is truncated to $Ro_{zon}\geq 10^{-3}$ to highlight the  supercritical cases as  $Ro_{zon}$ is very small near the onset.  In the oscillatory regime $Ro_c<0.2$, $Ro_{zon}$ also depends on the Ekman number, suggesting that the viscosity plays a part in the zonal flow. Based on a balance between the Reynolds stress and the internal viscous stress, \cite{Christensen2002} proposed 
$Ro_{zon} \propto \left(Ro_{non}\right)^2/E$.     
The non-zonal Rossby number as a function of $Ro_c$ is shown in figure \ref{fig:Rossby}(b), but the above relation does not hold for our numerical results as both $R_{zon}$ and $Ro_{non}$ follow the solid black line and roughly scale as $Ro_c^2$ in the oscillatory regime $Ro_c<0.2$. 

In the GT regime $0.2\le Ro_c\le1.5$, $Ro_{zon}$ becomes independent of the viscosity and the empirical scaling  $Ro_{zon}\sim Ro_c^{3/5}$ (black dashed line in figure \ref{fig:Rossby}(a))is evident, in agreement with {figure \ref{fig:ZonalProfilePr1}(b)}. In this regime, the non-zonal Rossby number is expected to follow the so-called inertial scaling, which is based on a force balance between Coriolis, inertia and Archimedean force, if the viscous effect is negligible \citep{Aubert2001}. The inertial scaling can be expressed as the following based on our definition of the control parameters (see Appendix \ref{app:inertial_scaling}):
\begin{equation}
    Ro_{non}\sim \left(\frac{E}{Pr}\right)^{2/5}\left(Ro_c\right)^{4/5}.
\end{equation}
In figure \ref{fig:Rossby}(b), $Ro_{non}$ is divided by $E^{2/5}$ ($Pr=1$) to compare with the inertial scaling (black dashed line). While the data points at different $E$ are well collapsed, we note that numerical results exhibit { a steeper trend (larger exponent)} than the inertial scaling. Previous numerical simulations in spherical shells also found steeper exponents \citep[e.g.][]{Gastine2016,Long2020} and the deviation is usually attributed to the non-negligible effects of the viscosity in numerical simulations.

\subsection{Heat transport} \label{sec:Nu}

\begin{figure}
\begin{center}
\includegraphics[width=0.48 \textwidth]{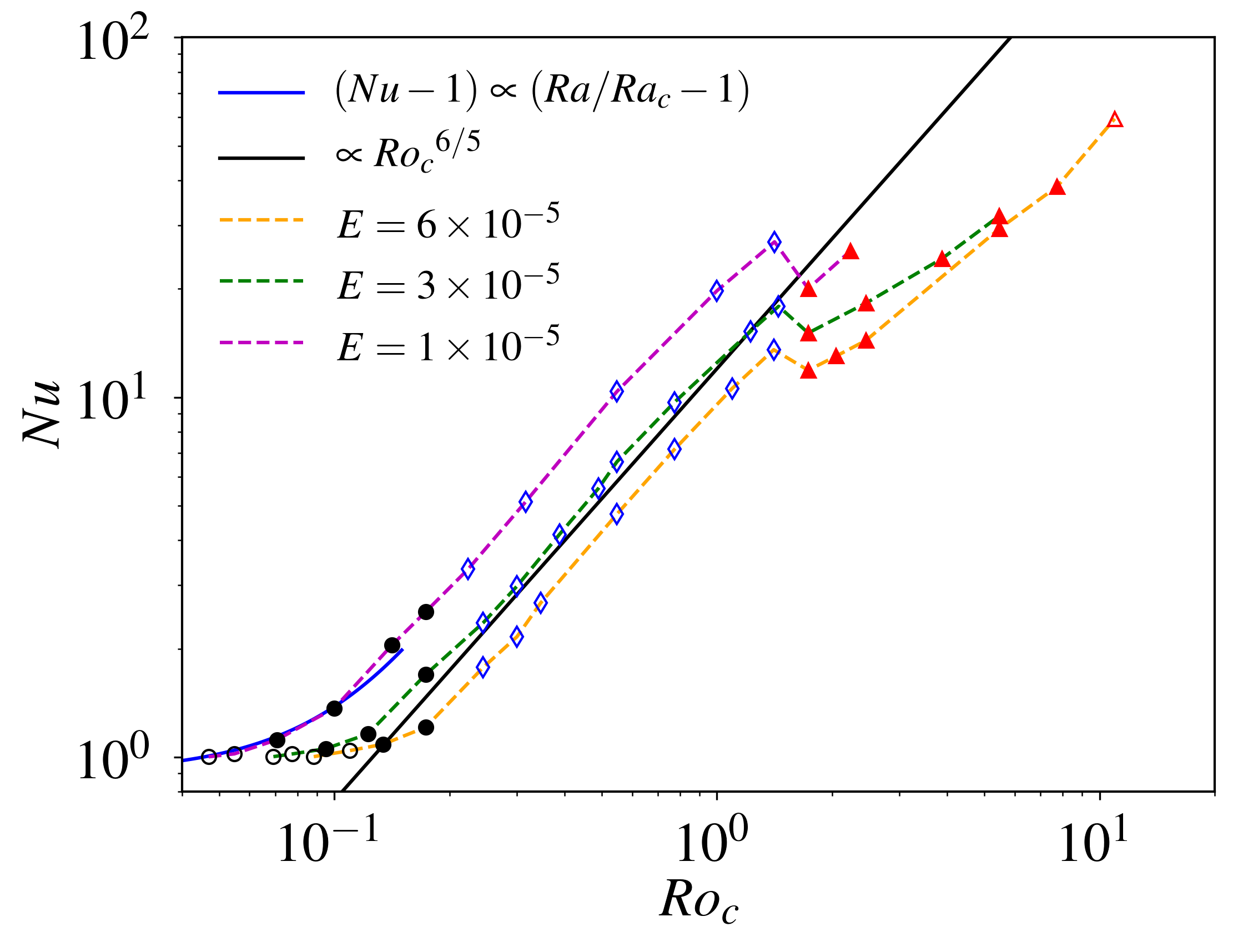}
\includegraphics[width=0.49 \textwidth]{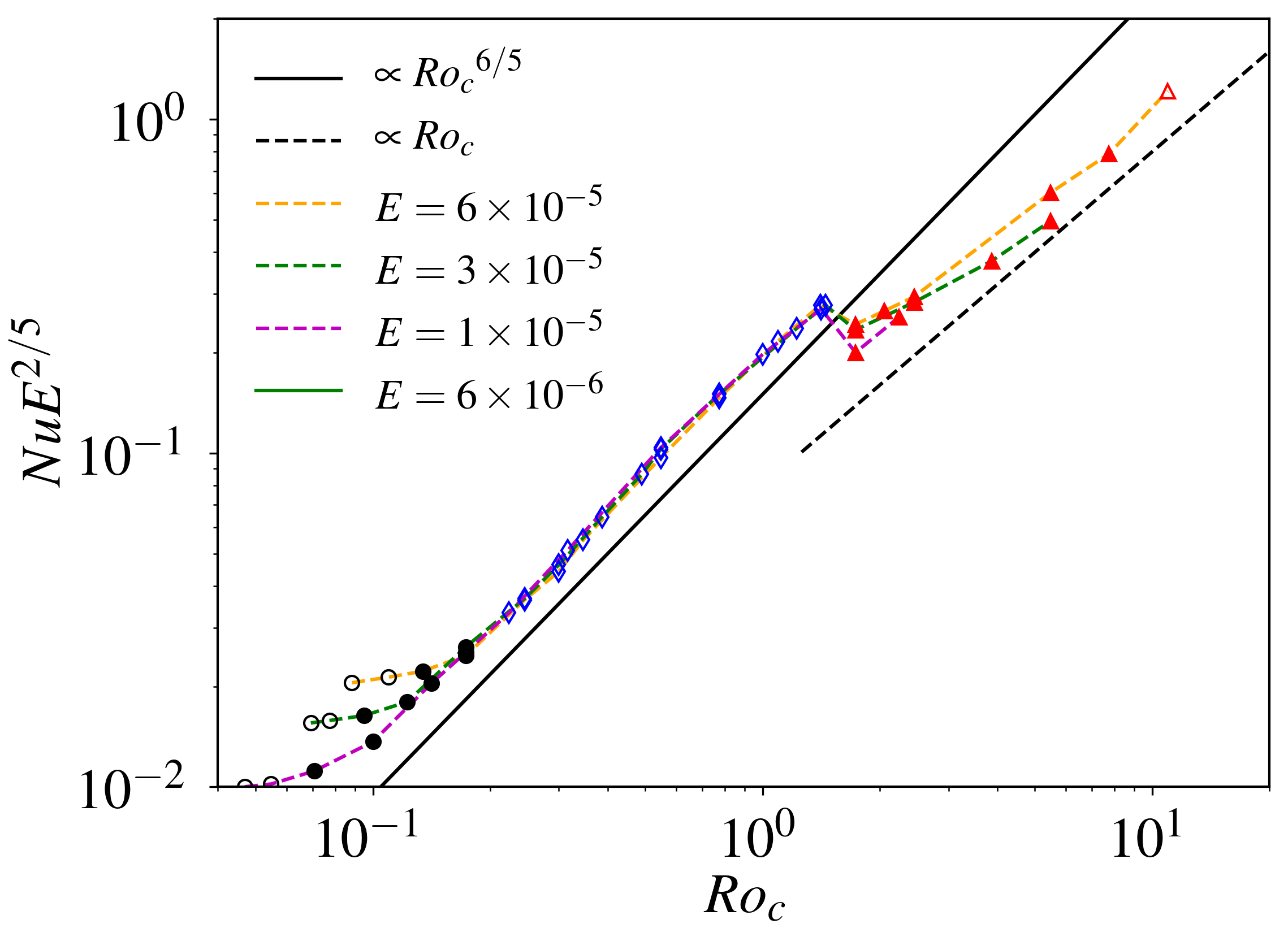} \\
(a)\hspace{6cm}(b)    
\caption{{Nusselt number $Nu$ (a) and $NuE^{2/5}$ (b) as a function of the convective Rossby number $Ro_c$. Different symbols represent different flow regimes as in figure \ref{fig:RegimDiagram}. The blue line depicting supercriticality is plotted in the lower left-hand corner of (a), for $Ro_c\le 0.15$ and $E=10^{-5}$. Here $Nu$ is time-averaged over several convective turnover time scales.}}
\label{fig:Nusselt}
\end{center}
\end{figure}

Figure \ref{fig:Nusselt} shows the Nusselt number as a function of the convective Rossby number $Ro_c$.
Near the onset of the convection, the convective heat transfer is proportional to the supercriticality according to a weakly nonlinear theory \citep{Gillet2006}, i.e.
\begin{equation}
    Nu-1 \propto \frac{Ra}{Ra_c}-1.
\end{equation}
We can see from the lower left-hand corner of figure \ref{fig:Nusselt}(a) that our numerical results can be well fitted by the aforementioned scaling when $Ra<6Ra_c$ for $E=10^{-5}$ (the blue solid line).

In rotating turbulent convection, the inertial scaling has been promoted to describe  heat transport {\citep{Stevenson1979, Aubert2001,Julien2012,Barker2014,Jones2015}}. Based on our definition of the non-dimensional parameters, this inertial scaling for the Nusselt number can be written as (Appendix \ref{app:inertial_scaling}):
{
\begin{equation}
    Nu \sim \left(\frac{E}{Pr}\right)^{-2/5}\left(Ro_c\right)^{6/5}.
\end{equation}}

We can see that our numerical results in the GT regime (blue diamonds in figure \ref{fig:Nusselt}) are in very good agreement with the inertial scaling. Note that we use the stress-free boundary condition in this study. Previous studies using the no-slip boundary found that the scaling exponent in $Ra$ depends on the Ekman number  \citep{Cheng2015,Gastine2016,Long2020}, i.e. $Nu\sim Ra^{\lambda(E)}$, though it may asymptotically approach the inertial scaling at low Ekman numbers \citep{Gastine2016}.   

There is a sudden drop of $Nu$ { on} increasing $Ro_c$ when the flow transits from the GT regime to the LSV regime $Ro_c>1.5$. As we have mentioned, the reduction of the heat transport by the LSV is mainly due to the fact that a relatively hot column is formed around the rotation axis. Nusselt number $Nu$ continues to increase on further increase of $Ra$ in the LSV regime, but the slope seems to be less steep compared to the GT regime. It is difficult to deduce any definitive scaling tendency, as higher forcings would be required, and this is computationally prohibitive.
\begin{figure}
\begin{center}
\includegraphics[height=0.25\textwidth]{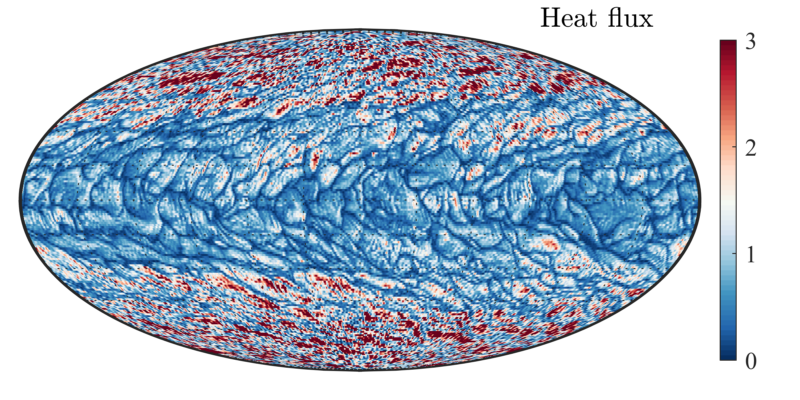}
\includegraphics[height=0.25\textwidth]{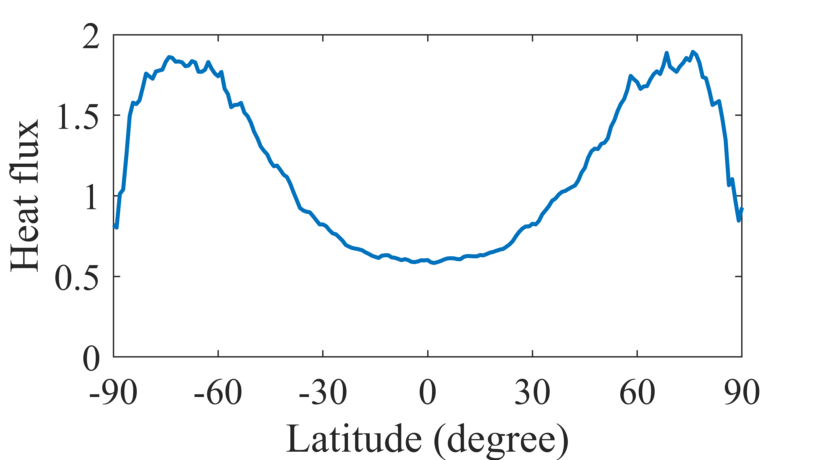}
\\
(a)\hspace{5cm}(b) \\
\caption{Heat flux $q$ on the surface in the LSV regime at $E=10^{-5}$ and $Ra=3.0\times 10^{5}$. (a) Instantaneous heat flux on the surface (Mollweide projection). (b) Time and azimuthal average of the heat flux $q$ on the surface as a function of latitude. }
\label{fig:HeatSurface}
\end{center}
\end{figure}

{ Figure \ref{fig:HeatSurface} (a) shows the heat flux $q$ on the surface in the LSV regime at $E=10^{-5}$ and $Ra=3.0\times 10^{5}$. The heat flux exhibits both small-scale perturbations and large-scale coherent variation in latitude, in line with the formation of LSV. The time and azimuthal average in figure \ref{fig:HeatSurface} (b) shows that the heat flux at high latitudes is about three times larger than that around the equator. The latitudal variations of the heat flux again suggest that the formation of LSV provides a barrier for the horizontal heat transport in the system.}

\subsection{Effects of boundary conditions}
Thus far, we have made use of  stress-free and fixed temperature boundary conditions. In this section, we briefly discuss effects of boundary conditions on the formation of LSV.

\begin{figure}
    \centering
    \includegraphics[width=0.7\textwidth]{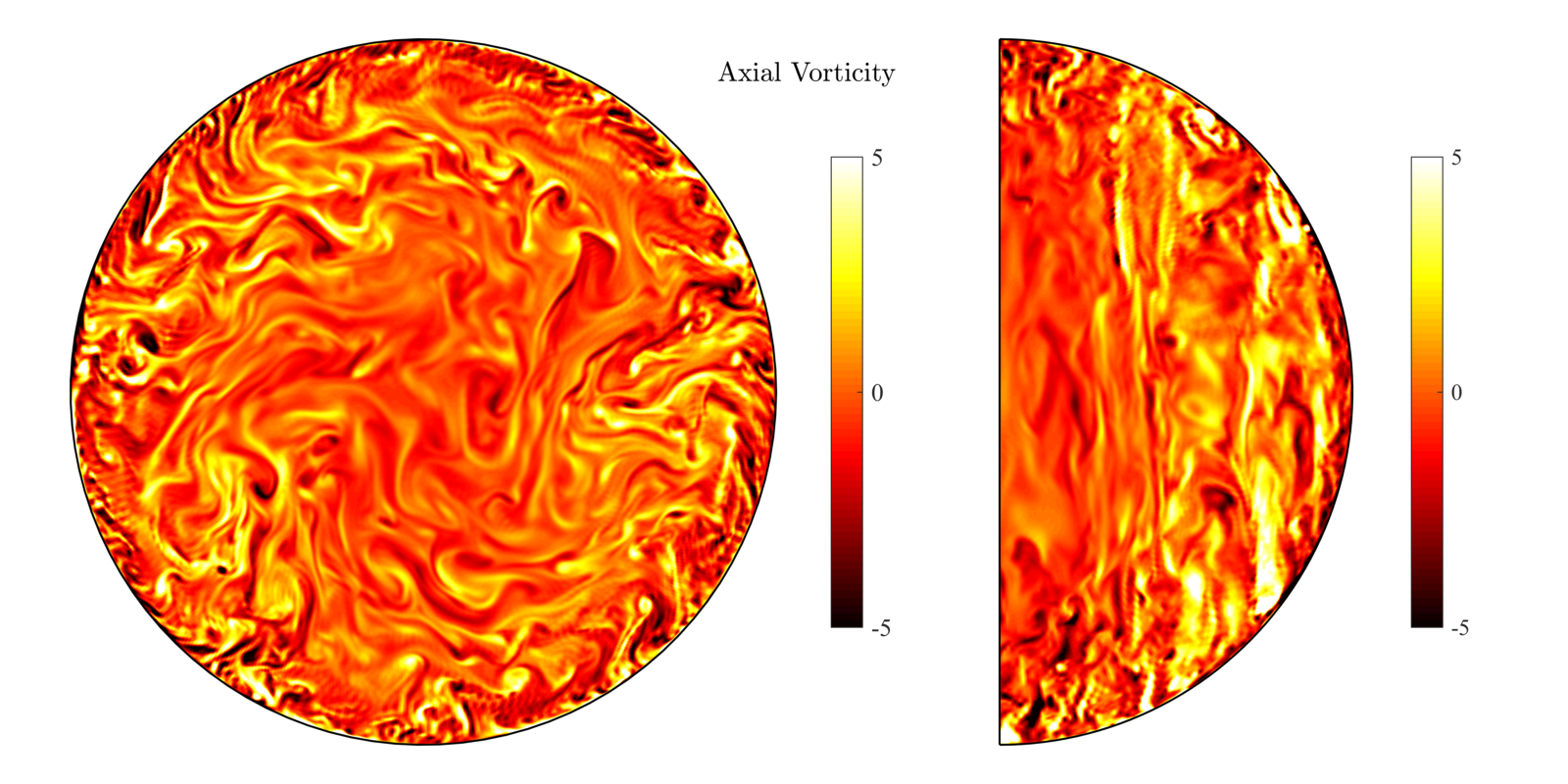}
    \includegraphics[width=0.25 \textwidth]{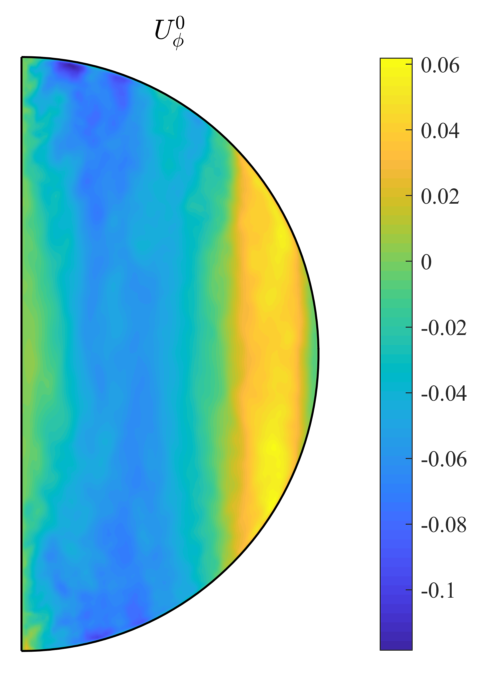}\\
    (a) \hspace*{3cm} (b)\hspace*{3cm} (c) \\
    \caption{Axial vorticity and zonal flow with the no-slip boundary condition at $E=1.0\times 10^{-5}$ and $Ra=5.0\times10^5$. (a-b) Axial vorticity in the equatorial plane and in the meridional plane; (c) zonal flow in the meridional plane.}
    \label{fig:Vor_NS}
\end{figure}

\begin{figure}
\begin{center}
\includegraphics[width=0.49 \textwidth]{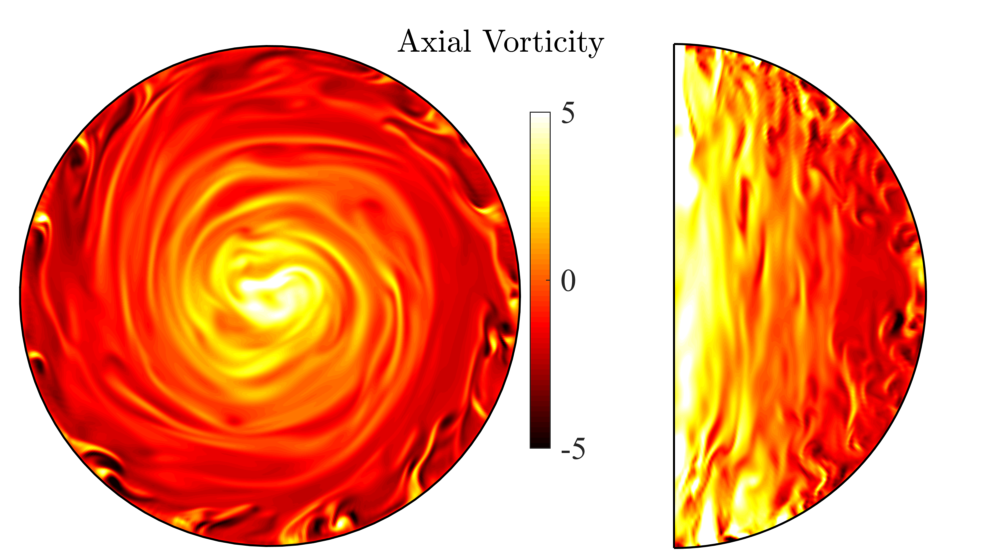}
\includegraphics[width=0.49 \textwidth]{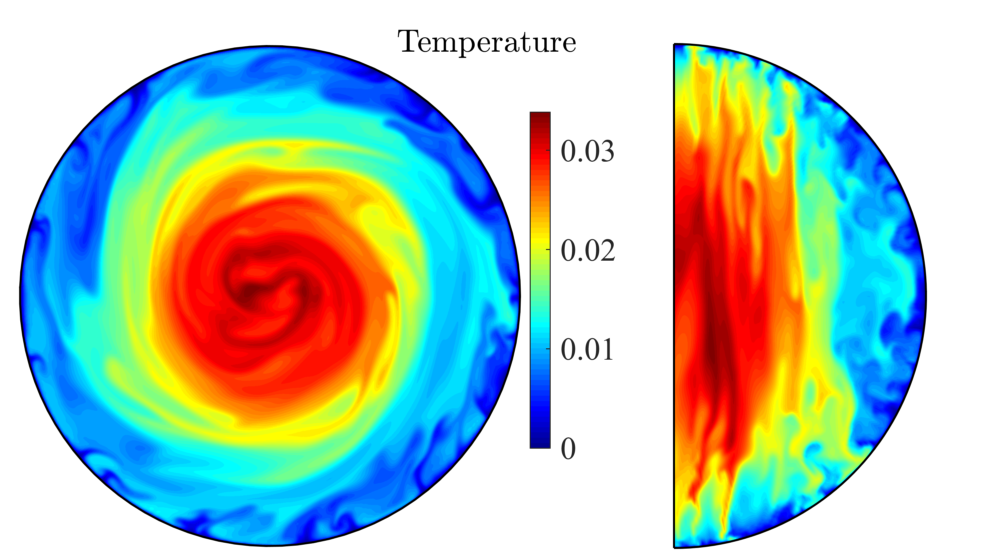} \\
(a)\hspace{5cm} (b) \\
\includegraphics[width=0.49 \textwidth]{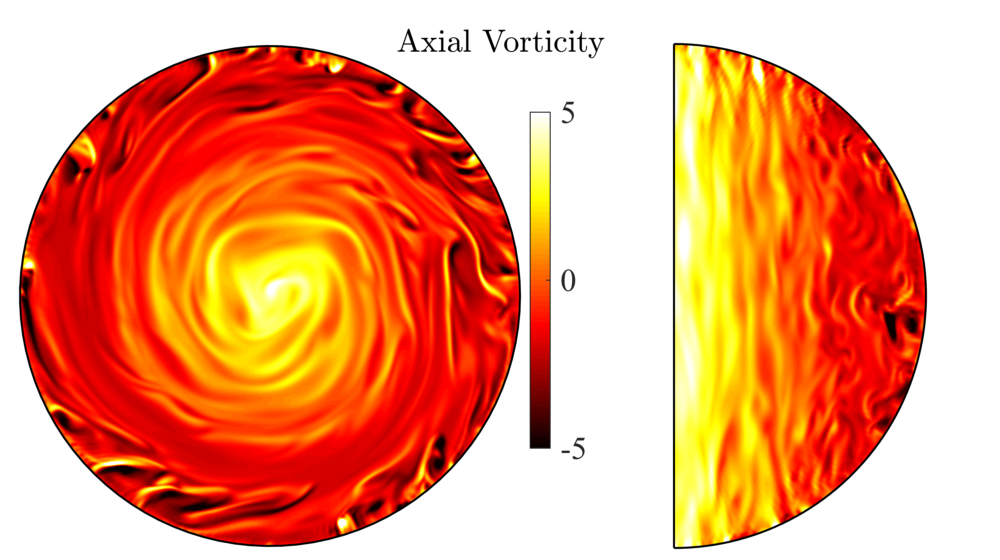} 
\includegraphics[width=0.49 \textwidth]{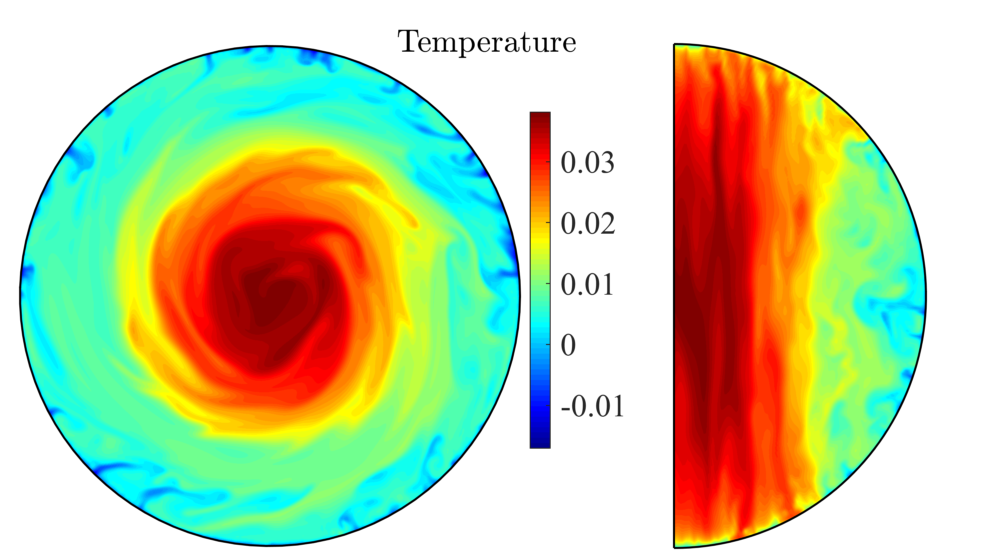} \\
(c)\hspace{5cm} (d)
\caption{Comparison of the fixed temperature boundary condition (a-b) and the fixed flux boundary condition (c-d) for $E=3.0\times 10^{-5}$ and $Ra=10^{5}$. (a) and (c) Axial vorticity in the equatorial plane and in the meridional plane. (b) and (d) Temperature in the equatorial plane and in the meridional plane.}
\label{fig:ThermalBCS}
\end{center}
\end{figure}

Previous studies {\citep{Guervilly2014,Stellmach2014}}, studying  planar rotating convection, found that the LSV exists only when a stress-free boundary condition is imposed. In the present study, we briefly explore the possibility of formation of LSV with a no-slip boundary, but do not observe any  LSV being formed, at least in the parameter range we could access. Figure \ref{fig:Vor_NS} shows an example with a no-slip boundary condition at $E=10^{-5}$ and $Ra=5\times 10^{5}$, at which the LSV would form with the stress-free boundary condition. With the no-slip boundary condition, we see that the flow is {characterised} by small-scale structures elongated along the rotation axis (figure \ref{fig:Vor_NS}(a-b)). We note that a zonal flow is also developed with a prograde jet near the equator and a retrograde flow in the inner region (figure \ref{fig:Vor_NS}(c)), but the amplitude is much smaller compared to the zonal flow with a stress-free boundary condition. In fact, the zonal Rossby number is smaller than the non-zonal Rossby number when  no-slip boundary conditions are employed (table \ref{tab:summary}), in contrast to the case with the stress-free boundary condition. 

Figure \ref{fig:ThermalBCS} compares different thermal boundary conditions, i.e.  fixed temperature (\ref{fig:ThermalBCS}a-b) and  fixed flux (\ref{fig:ThermalBCS}c-d) conditions at $E=1.0\times 10^{-5}$ and $Ra=5.0\times10^5$. Both cases use the stress-free boundary condition. It can be seen that in this regime the thermal boundary conditions play a minor role. A major difference  is the temperature field in the polar regions, where the hot columnar structure associated with the LSV needs to adapt different thermal boundary conditions. The hot columnar region can extend to the polar surface when the fixed-flux condition is imposed.

\section{Conclusions}\label{sec:Conclusion}

We have observed the previously unseen presence of LSVs in full-sphere rotating convection, thus complementing the previous discoveries in Boussinesq rotating boxes
  \citep{Favier2014,Guervilly2014,Stellmach2014,Rubio2014}. Relatively simple large-scale zonal flow profiles are developed in this regime, with a prograde axial vortex and a retrograde equatorial zonal flow. While the geostrophic character
of these flows is pervasive, the ubiquitous radial structure of the zonal flow  in the LSV regime remains a 
theoretical challenge to comprehend. We have been able to fully document the transition between regimes as the Rayleigh number is increased from that required to first initiate convection, finding that the convective Rossby number is a completely prognostic control parameter: $Ro_c\approx 0.2$ and $Ro_c\approx1.5$ delineate the boundaries between the oscillatory regime, the geostrophic turbulence  regime and the LSV regime. 

In the {geostrophic turbulence} regime, the zonal Rossby number $Ro_{zon}$ solely depends on the convective Rossby number and follows an empirical scaling law $Ro_{zon}\sim Ro_c^{3/5}$, while the non-zonal Rossby number and the Nusselt number can be described by the well-known inertial scaling. The direction of the zonal flow is reversed when the flow transitions from the {geostrophic turbulence} regime to the LSV regime, while the non-zonal Rossby number and the Nusselt number are reduced by the formation of LSV.
In the LSV regime we witness a saturation of the zonal flow speed, while the non-zonal Rossby number and the Nusselt number show an abrupt drop followed by a gradual increase, whose slopes are reduced compared to their systematics in the {geostrophic turbulence} regime.

The effects of both the mechanical and thermal boundary conditions on the formation of LSVs are briefly considered. While the thermal boundary condition plays a minor role, the no-slip boundary condition prohibits the formation of LSVs, at least for the parameter regimes we could access, in line with previous studies in planar rotating convection \citep{Guervilly2014,Stellmach2014}. It is possible that the dynamics with the stress-free and the no-slip conditions would converge at very low Ekman numbers, but it is computationally demanding to study the highly supercrtical convection at very low Ekman numbers via direct numerical simulations.
We note the prospect of better experimental delineation of zonal flows in the 
future using the ZoRo experiment \citep{SU2020}. This rapidly rotating spheroid has the capability of zonal flow determination using acoustic normal mode velocimetry, and will be subject to differential heating in the future.

\section*{Declaration of Interests} 
The authors report no conflict of interest.

\section*{Acknowledgements}
We thank Philippe Marti for the provision of the computer code EPMDynamo, and much implementation advice. 
Simulations were performed  at the Swiss National Supercomputer Centre (CSCS) under account s872, on the Euler cluster of ETH Zurich and on the Taiyi cluster supported by the Center for Computational Science and Engineering of Southern University of Science and Technology. The research was partially supported by a European Research Council Advanced grant 833848 (UEMHP) to AJ under the European Union’s Horizon 2020 research and innovation programme. YL was supported by the B-type Strategic Priority Program of the Chinese Academy of Sciences
(XDB41000000), the National Natural Science Foundation of China (grant No.41904066), the pre-research project on Civil Aerospace Technologies of China National Space Administration (D020308) and the Macau Foundation. YL was partly supported by a Swiss NSF Advanced PostDoc Mobility Fellowship when this study was initiated. 

\appendix
\section{List of numerical simulations}\label{appA}
All of the numerical simulations are listed in table \ref{tab:summary} including the control parameters, time-averaged diagnostic parameters, numerical resolutions ($N,L,M$) and flow regimes. {Here SD represents the steadily drifting regime near the onset; RO represents the relaxation oscillation regime; GT represents the geostrophic turbulence regime;} LSV represents the large scale vortices; NR represents the non-rotating regime. The case marked with \# used the stress-free and fixed flux boundary condition, while the one marked with * used the no-slip and fixed temperature boundary conditions. All other cases used the stress-free and fixed temperature boundary conditions.
\begin{table}
 \begin{center}
\def~{\hphantom{0}}
  \begin{tabular}{cccccccp{.1\textwidth}} 
    $E$   & $Ra$  & $Ro_{zon}$ & $Ro_{non}$& $Nu$ & $(N,L,M)$ & Regime \\[5pt]
   $6.0\times 10^{-5}$  & $Ra_c=126.04$ & \\ 
   $6.0\times 10^{-5}$  & $1.3\times 10^2$& $2.85\times 10^{-5}$ &$1.55\times10^{-4} $& 1.002  &(63,63,63)& SD \\
   $6.0\times 10^{-5}$  & $2.0\times 10^{2}$& $5.81\times 10^{-4}$ &$6.94\times10^{-4} $& 1.041  &(63,63,63)& SD \\
   $6.0\times 10^{-5}$  & $3.0\times 10^2$ & $1.56\times 10^{-3}$  & $1.20\times 10^{-3}$ & 1.085  &(63,63,63)& RO \\
   $6.0\times 10^{-5}$  & $5.0 \times 10^2$ & $3.77\times 10^{-3}$  &$2.02\times 10^{-3}$ &1.774  &(63,63,63)& RO \\
   $6.0\times 10^{-5}$  & $1.0\times 10^3$ & $9.08\times 10^{-3}$  & $3.93\times 10^{-3}$ & 1.774  &(63,63,63)& GT \\
   $6.0\times 10^{-5}$  & $1.5\times 10^3$ & $1.26\times 10^{-2}$  & $4.90\times 10^{-3}$ & 2.158  &(63,63,63)& GT \\
   $6.0\times 10^{-5}$  & $2.0\times 10^3$ & $1.49\times 10^{-2}$  & $5.80\times 10^{-3}$ & 2.683  &(63,63,63)& GT \\
    $6.0\times 10^{-5}$ & $5.0\times 10^3$ & $2.23\times 10^{-2}$  & $1.07\times 10^{-2}$ & 4.731  &(63,127,127)& GT \\
    $6.0\times 10^{-5}$ & $1.0\times 10^4$ & $2.87\times 10^{-2}$  & $2.27\times 10^{-2}$ & 7.162  &(63,127,127)& GT \\
    $6.0\times 10^{-5}$ & $2.0\times 10^4$ & $3.65\times 10^{-2}$  & $3.15\times 10^{-2}$ & 10.57  &(63,127,127)& GT \\
    $6.0\times 10^{-5}$ & $3.3\times 10^4$ & $4.11\times 10^{-2}$  & $3.99\times 10^{-2}$ & 13.55  &(63,127,127)& GT \\
    $6.0\times 10^{-5}$ & $5.0\times 10^4$ & $1.05\times 10^{-1}$  & $3.80\times 10^{-2}$ & 11.88  &(63,127,127)& LSV \\
    $6.0\times 10^{-5}$ & $7.0\times 10^4$ & $1.20\times 10^{-1}$  & $4.37\times 10^{-2}$ & 13.01  &(63,127,127)& LSV \\
    $6.0\times 10^{-5}$ & $1.0\times 10^5$ & $1.33\times 10^{-1}$  & $6.06\times 10^{-2}$ & 14.37  &(63,127,127)& LSV \\
    $6.0\times 10^{-5}$ & $5.0\times 10^5$ & $1.54\times 10^{-1}$  & $1.26\times 10^{-1}$ & 29.43  &(127,255,255)& LSV \\
    $6.0\times 10^{-5}$ & $1.0\times 10^6$ & $1.50\times 10^{-1}$  & $1.32\times 10^{-1}$ & 38.49  &(127,255,255)& LSV \\
    $6.0\times 10^{-5}$ & $2.0\times 10^6$ & $1.46\times 10^{-1}$  & $1.81\times 10^{-1}$ & 59.41  &(127,255,255)& NR \\
    [5pt]
 $3.0\times 10^{-5}$  & $Ra_c=152.95$ & \\
 $3.0\times 10^{-5}$  & $1.6\times 10^2$ & $2.43\times 10^{-5}$ &$9.62\times10^{-5}$ &1.002  &(63,63,63)& SD \\
 $3.0\times 10^{-5}$  & $2.0\times 10^2$ & $2.06\times 10^{-4}$ &$2.93\times10^{-4}$ &1.019  &(63,63,63)& SD \\
 $3.0\times 10^{-5}$  & $3.0\times 10^2$ & $7.54\times 10^{-4}$ &$5.36\times10^{-4}$ &1.054  &(63,63,63)& RO \\
 $3.0\times 10^{-5}$  & $5.0\times 10^2$ & $2.09\times 10^{-3}$ &$9.18\times10^{-4}$ &1.159  &(63,127,127)& RO \\
 $3.0\times 10^{-5}$  & $1.0\times 10^3$ & $5.83\times 10^{-3}$ &$2.22\times10^{-3}$ &1.694  &(63,127,127)& RO \\
 $3.0\times 10^{-5}$  & $2.0\times 10^3$ & $1.10\times 10^{-2}$ &$3.38\times10^{-3}$ &2.361  &(63,127,127)& GT \\
 $3.0\times 10^{-5}$  & $3.0\times 10^3$ & $1.38\times 10^{-2}$ &$4.23\times10^{-3}$ &2.982 &(63,127,127)& GT \\
 $3.0\times 10^{-5}$  & $5.0\times 10^3$ & $1.71\times 10^{-2}$ &$5.84\times10^{-3}$ &4.152  &(63,127,127)& GT \\
 $3.0\times 10^{-5}$  & $8.0\times 10^3$ & $2.04\times 10^{-2}$ &$7.75\times10^{-3}$ &5.577  &(63,127,127)& GT \\
 $3.0\times 10^{-5}$  & $1.0\times 10^4$ & $2.21\times 10^{-2}$ &$8.79\times10^{-3}$ &6.615  &(63,127,127)& GT \\
 $3.0\times 10^{-5}$  & $2.0\times 10^4$ & $2.86\times 10^{-2}$ &$1.33\times10^{-2}$ &9.684  &(63,127,127)& GT \\
 $3.0\times 10^{-5}$  & $5.0\times 10^4$ & $3.78\times 10^{-2}$ &$2.93\times10^{-2}$ &15.25  &(63,127,127)& GT \\
 $3.0\times 10^{-5}$  & $7.0\times 10^4$ & $3.92\times 10^{-2}$ &$3.56\times10^{-2}$ &17.89  &(63,127,127)& GT \\
 $3.0\times 10^{-5}$  & $1.0\times 10^5$ & $1.04\times 10^{-1}$ &$2.46\times10^{-2}$ &15.08  &(127,255,255)& LSV\\
 $3.0\times 10^{-5}$  & $1.0\times 10^5$ & $1.05\times 10^{-1}$ &$2.58\times10^{-2}$ &--  &(127,255,255)& LSV $^\#$\\
 $3.0\times 10^{-5}$  & $2.0\times 10^5$ & $1.23\times 10^{-1}$ &$3.76\times10^{-2}$ &18.25  &(127,255,255)& LSV\\
 $3.0\times 10^{-5}$  & $5.0\times 10^5$ & $1.54\times 10^{-1}$ &$8.13\times10^{-2}$ &24.27  &(127,255,255)& LSV\\
  $3.0\times 10^{-5}$  & $1.0\times 10^6$ & $1.67\times 10^{-1}$ &$9.65\times10^{-2}$ &31.92  &(127,255,255)& LSV\\
 [5pt]     
 $1.0\times 10^{-5}$  & $Ra_c=210.52$ & \\
 $1.0\times 10^{-5}$  & $2.2\times10^2$& $1.01\times10^{-5}$ & $3.42\times10^{-5}$ &1.001  &(63,63,63)& SD \\
 $1.0\times 10^{-5}$  & $3.0\times 10^2$ & $1.75\times10^{-4}$ & $1.52\times10^{-4}$ &1.020  &(63,127,127)& SD \\
 $1.0\times 10^{-5}$  & $5.0\times 10^2$ & $7.25\times10^{-4}$ & $2.17\times10^{-4}$ &1.116  &(63,127,127)& RO\\
$1.0\times 10^{-5}$  &  $1.0\times 10^3$ & $2.56\times10^{-3}$ & $4.52\times10^{-4}$ &1.366  &(63,127,127)& RO\\
$1.0\times 10^{-5}$  &  $2.0\times 10^3$ & $6.19\times10^{-3}$ & $1.17\times10^{-3}$ &2.046  &(63,127,127)& RO\\
$1.0\times 10^{-5}$  &  $5.0\times 10^3$ & $1.11\times10^{-2}$ & $2.25\times10^{-3}$ &3.326  &(63,127,127)& RO\\
 $1.0\times 10^{-5}$  & $1.0\times 10^4$ & $1.44\times10^{-2}$ & $3.46\times10^{-3}$ &5.120  &(127,255,255)& GT\\
 $1.0\times 10^{-5}$  & $3.0\times 10^4$ & $2.12\times10^{-2}$ & $6.60\times10^{-3}$ &10.37  &(127,255,255)& GT\\
 $1.0\times 10^{-5}$  & $3.0\times 10^4$ & $2.12\times10^{-2}$ & $6.60\times10^{-3}$ &10.37  &(127,255,255)& GT\\
 $1.0\times 10^{-5}$  & $1.0\times 10^5$ & $3.23\times10^{-2}$ & $1.18\times10^{-2}$ &19.76  &(127,255,255)& GT\\
  $1.0\times 10^{-5}$  & $2.0\times 10^5$ & $3.84\times10^{-2}$ & $1.74\times10^{-2}$ &27.03  &(127,255,255)& GT\\
  $1.0\times 10^{-5}$  & $3.0\times 10^5$ & $1.08\times10^{-1}$ & $1.56\times10^{-2}$ &20.03  &(127,255,255)& LSV\\
  $1.0\times 10^{-5}$  & $5.0\times 10^5$ & $1.19\times10^{-1}$ & $1.92\times10^{-2}$ &29.57  &(127,255,255)& LSV\\
  $1.0\times 10^{-5}$  & $5.0\times 10^5$ & $2.14\times 10^{-2}$& $2.60\times10^{-2}$ &43.17 & (127,255,255)& GT *\\
 \end{tabular}
  \caption{Summary of numerical simulations. Symbols $^\#$ and * denote the cases using different boundary boundary conditions (see text for details).}
  \label{tab:summary}
  \end{center}
\end{table}

\section{Inertial scaling expressed in our parameters} \label{app:inertial_scaling}
In rapidly rotating convection, the so-called inertial scaling is proposed to describe the convective flow speed, typical length scale and heat transport \citep{Stevenson1979,Aubert2001}. The inertial scaling is based on a balance of Coriolis-Inertial-Archimedean (CIA) terms, incorporated with the mixing length theory, leading to a scaling independent of the viscosity. Here we follow the review article of \cite{Jones2015}  and  express the scaling in terms of dimensionless parameters defined in this study. Note that the Rayleigh number $Ra$ we defined is rotationally modified and heat-flux based, which is loosely related to the $Ra_Q$ in \cite{Jones2015} by $Ra\approx ERa_Q$. According to equation (119) in \cite{Jones2015}, the inertial scaling predicts the non-zonal Rossby number (representing the convective flow velocity) as 
\begin{equation}
    Ro_{non}\sim \left(\frac{E}{Pr}\right)(EPr)^{1/5}Ra_Q^{2/5} \sim \left(\frac{E}{Pr}\right)^{4/5}Ra^{2/5}.
\end{equation}
We can write the above scaling in the convective Rossby number $Ro_c=\sqrt{RaE/Pr}$ as
\begin{equation}
    Ro_{non}\sim \left(\frac{E}{Pr}\right)^{2/5}Ro_c^{4/5}.
\end{equation}
Note that the inertial scaling does not provide direct prediction of the zonal Rossby number.

The inertial scaling predicts the Nusselt number as \citep[equation (119) in][]{Jones2015}
\begin{equation}
    Nu \sim E^{4/5}Pr^{-1/5}Ra_Q^{3/5} \sim \left(\frac{E}{Pr}\right)^{1/5}Ra^{3/5}.
\end{equation}
Note that we have assumed $Nu-1 \sim Nu$ in the strongly supercritical regime. Using $Ra^*=RaE/Pr$, we obtain
\begin{equation}\label{eq:NuRa}
    Nu \sim \left(\frac{E}{Pr}\right)^{-2/5}\left(Ra^*\right)^{3/5}.
\end{equation}
The above scaling is equivalent to
\begin{equation}
    Nu \sim E^2 Pr^{-1/2} Ra_T^{3/2},
\end{equation}
where $Ra_T$ is the convectional Rayleigh number based on the temperature difference, i.e. $Ra_{T}=\alpha g_0 \Delta T r_o^3/{\nu \kappa}$, and is related to $Ra$ by 
\begin{equation}
    Ra=2NuERa_{T}.
\end{equation}
{
Equation (\ref{eq:NuRa}) can be also represented in terms of the convective Rossby number $Ro_c$ using $Ra^*=Ro_c^2$:
\begin{equation}
    Nu \sim \left(\frac{E}{Pr}\right)^{-2/5}\left(Ro_c\right)^{6/5}.
\end{equation}
}

\bibliographystyle{jfm}
\bibliography{reference}

\end{document}